\documentclass[twocolumn]{pasj01}
\RequirePackage{mathpazo}
\RequirePackage[T1]{fontenc}

\usepackage{color}
\usepackage{comment}
\usepackage{lineno}

\def\commenta{$^*$}

\newcounter{author}
\def\authorcount#1#2{\refstepcounter{author}\label{#1}
                     \altaffiltext{\ref{#1}}{#2}}

\Received{$\langle$reception date$\rangle$}
\Accepted{$\langle$acception date$\rangle$}
\Published{$\langle$publication date$\rangle$}
\begin{document}
\SetRunningHead{Y. Tampo et al.}{spectra of V455 And superoutburst}

\title{Spectroscopic Observations of  V455 Andromedae Superoutburst in 2007: the Most Exotic Spectral Features in Dwarf Nova Outbursts}

\author{
        Yusuke~\textsc{Tampo}\altaffilmark{\ref{affil:Kyoto}*},
        Daisaku~\textsc{Nogami}\altaffilmark{\ref{affil:Kyoto}},
        Taichi~\textsc{Kato}\altaffilmark{\ref{affil:Kyoto}},
        Kazuya~\textsc{Ayani}\altaffilmark{\ref{affil:BAO}},
        Hiroyuki~\textsc{Naito}\altaffilmark{ \ref{affil:Nayoro}, \ref{affil:NHAO}}, 
        Norio~\textsc{Narita}\altaffilmark{\ref{affil:UTokyo}, \ref{affil:ABC}, \ref{affil:IAC_spain}},
        Mitsugu~\textsc{Fujii}\altaffilmark{\ref{affil:FKO}},        
        Osamu~\textsc{Hashimoto}\altaffilmark{\ref{affil:Gunma}},        
        Kenzo~\textsc{Kinugasa}\altaffilmark{\ref{affil:Nobeyama}, \ref{affil:Gunma}},        
        Satoshi~\textsc{Honda}\altaffilmark{\ref{affil:NHAO}, \ref{affil:Gunma}},        
        Hidenori~\textsc{Takahashi}\altaffilmark{\ref{affil:kiso}, \ref{affil:Gunma}},        
        Shin-ya~\textsc{Narusawa}\altaffilmark{\ref{affil:NHAO}},
        Makoto~\textsc{Sakamoto}\altaffilmark{\ref{affil:NK}, \ref{affil:NHAO}},
        and Akira~\textsc{Imada}\altaffilmark{\ref{affil:Kyoto}}
}

\authorcount{affil:Kyoto}{
     Department of Astronomy, Kyoto University, Kitashirakawa-Oiwake-cho, Sakyo-ku, 
     Kyoto, 606-8502. Japan}
\email{tampo@kusastro.kyoto-u.ac.jp}

\authorcount{affil:BAO}{
    Bisei Astronomical Observatory, 1723-70 Okura, Bisei, Ibara,
    Okayama 714-1411, Japan}

\authorcount{affil:Nayoro}{
    Nayoro Observatory, 157-1 Nisshin, Nayoro, Hokkaido 096-0066, Japan}

\authorcount{affil:NHAO}{
    Nishi-Harima Astronomical Observatory, Center for Astronomy,
    University of Hyogo, 407-2 Nishigaichi, Sayo-cho, Sayo, Hyogo
    679-5313, Japan}

\authorcount{affil:UTokyo}{
    Komaba Institute for Science, The University of Tokyo, 3-8-1 Komaba, Meguro, Tokyo 153-8902, Japan}

\authorcount{affil:ABC}{
    Astrobiology Center, National Institutes of Natural Sciences, 
    2-21-1 Osawa, Mitaka, Tokyo 181-8588, Japan}
    
\authorcount{affil:IAC_spain}{
    Instituto de Astrof\'isica de Canarias, V\'ia L\'actea s/n,
    E-38205 La Laguna, Tenerife, Spain}

\authorcount{affil:FKO}{
    Fujii Kurosaki Observatory, 4500 Kurosaki, Tamashima, Kurashiki, Okayama
    713-8126, Japan}

\authorcount{affil:Gunma}{
    Gunma Astronomical Observatory, 6860-86 Nakayama, Takayama,
    Agatsuma, Gunma 377-0702, Japan}
    
\authorcount{affil:Nobeyama}{
    Nobeyama Radio Observatory, National Astronomical Observatory of Japan, 
    National Institutes of Natural Sciences, 462-2 Nobeyama, Minamimaki, 
    Minamisaku, Nagano 384-1305}

\authorcount{affil:kiso}{
    Kiso Observatory, Institute of Astronomy, School of Science,
    The University of Tokyo,
    10762-30 Mitake, Kiso-machi, Kiso-gun, Nagano 397-0101, Japan}

\authorcount{affil:NK}{
    Nenohoshi Kyoikusya, 231-5 Kajiya, Shingucho, Tatsuno, Hyogo 679-5154, Japan}

%%% end:list of authors


\maketitle

\begin{abstract}

We present our spectroscopic observations of V455 Andromedae during the 2007 superoutburst.
Our observations cover this superoutburst from around the optical peak of the outburst to the post-superoutburst stage.
During the early superhump phase, the emission lines of Balmer series, He I, He II, Bowen blend, and C IV / N IV blend  were detected.
He II 4686 line exhibited a double-peaked emission profile, 
where Balmer emission lines were single-peaked, 
which is unexpected from its high inclination.
In the ordinary superhump phase, Balmer series transitioned to double-peaked emission profiles, and high-ionization lines were significantly weakened.
These transitions of the line profiles should be related to the structural transformation of the accretion disk, as expected between the early and ordinary superhump transition in the thermal-tidal instability model.
The Doppler map of H$\alpha$ during the early superhump phase exhibits a compact blob centered at the primary white dwarf.
In analogy to SW Sex-type cataclysmic variables, this feature could  emerge from the disk wind and/or the mass accretion column onto the magnetized white dwarf. 
The Doppler map of He II 4686 \AA~ is dominated by the ring-like structure and imposed two flaring regions with the velocity of $\sim$300 km/s, which is too slow for a Keplerian accretion disk.
The phase of the flaring regions was coincident with the inner spiral arm structure identified during the early superhump phase.
Our disk wind model with the enhanced emission from the inner arm structure successfully reproduced the observed properties of He II 4686 \AA.
Therefore, V455 And is the first case in dwarf nova outbursts that the presence of the disk wind is inferred from an optical spectrum.

\end{abstract}

%\tableofcontents
%\linenumbers

\section{Introduction}
\label{sec:1}

Cataclysmic variables (CVs) are close binary systems composed of a primary white dwarf (WD) and a Roche lobe-filling secondary low-mass star  (for a review of CV, see \cite{war95book, hel01book}). 
Dwarf nova (DN) is a subclass of CVs which possesses an accretion disk and exhibits recurrent outbursts triggered by the thermal-tidal instability \citep{osa96review}.
WZ Sge-type DNe are the most evolved class of DNe, and they show only large-amplitude and long-duration superoutbursts every 5 - 30 yr (for a review, see \cite{kat15wzsge}). 
Outbursts of WZ Sge-type DNe are characterised by both early superhumps and ordinary superhumps.
Early superhumps are believed to be the result of double spiral structure in the disk that emerges from the 2:1 resonance \citep{lin79lowqdisk, osa02wzsgehump}, whereas ordinary superhumps are induced by the eccentric disk through the 3:1 resonance \citep{whi88tidal, osa89suuma}.
Early superhumps are a double-peaked variation with the period almost identical with the orbital period of the system \citep{ish02wzsgeletter,kat02wzsgeESH}.
Since a larger amplitude of early superhumps is observed in a system with higher inclination, the origin of early superhumps is believed to be the vertical deformation of the accretion disk \citep{nog97alcom, mae07bcuma, mat09v455and, uem12ESHrecon}.
However, observationally, the exact system structures of the spiral arms during the early superhump phase are still not resolved.
\citet{bab02wzsgeletter, kuu02wzsge} studied the Doppler tomography of He II 4686 \AA~ during the early superhump phase of WZ Sge.
They revealed an asymmetric spiral structure in the accretion disk, which can be attributed to the 2:1 resonance.

Another important viewpoint on WZ Sge-type outburst is the evolution of the disk structure across the superhump stages.
Since the 2:1 resonance suppresses the growth of the 3:1 resonance \citep{lub91SHa}, early superhump are always observed before the ordinary superhumps.
The evolution of early and ordinary superhumps are  well established in time-resolved photometric studies (e.g., \cite{Pdot} and following series).
Moreover, \citet{neu18j1222gwlib} found that the X-ray flux in DN outbursts is correlated with the superhump stages, and \citet{nii21a18ey} showed that the X-ray hard-soft transition in low-mass black hole X-ray binary ASASSN-18ey (=MAXI J1820+070) occurred simultaneously with the transition of optical superhump stages.
Although the spectra during the outbursts of WZ Sge-type DNe dramatically changes as well (e.g., WZ Sge: \cite{nog04wzsgespec}, GW Lib: \cite{hir09gwlib}), an association between the evolution of superhumps  and the spectra has been less discussed.
\citet{nog04wzsgespec, hir09gwlib} revealed that before the outburst peaks, their spectra were dominated by Balmer and He I absorption lines, whereas after the maxima they turned into emission lines and high excitation lines such as He II and C III / N III Bowen blend lines appeared.
These high excitation lines are then decayed out as the outburst proceeds, due to the decrease of irradiating UV photons and/or the disappearance of the spiral structure originating from the 2:1 resonance. 

While most of the accreting objects accompany an outflow component in  forms of a jet and a wind, CVs and DN outbursts have been believed that their outflow  is negligible in  optical studies.
Indeed,  optical spectra of DN outbursts are dominated by an emission from an accretion disk \citep{hor85zcha}.
Even though P Cygni profiles are observed in UV in some nova-like CVs (e.g., \cite{pri00bzcamwindHST}) and DN outbursts (e.g., \cite{szk99v1159ori,fro02diskwindDNproc,san11CVUV}),  only a handful nova-like CVs show P Cygni profile in optical (\cite{kaf04windfromCV} and reference therein) and never detected in the optical studies of DN outbursts in our best knowledge.
Another possible implication of an outflow component in CVs is single-peaked Balmer and He emission lines in eclipsing nova-like CVs, called SW Sex-type CVs \citep{hon86swsex}.
The theoretical studies by \citet{hon86swsex, mat15CVdiskwind} showed that these peculiar single-peaked emission lines from the eclipsing systems can originate from the wind component.
Recent studies have detected a radio emission during the outburst of firstly SS Cyg  \citep{kor08SSCyginradio} and then other systems \citep{cop16DNinRadio}.
The suggested emission mechanisms include a synchrotron emission from the transient jet, while the discussion does not settled.

V455 And (= HS 2331+3905) was originally found by the Hamburg Quasar Survey \citep{HQS} as a CV candidate.
\citet{ara05v455and} studied its quiescence photometrically and spectroscopically, showing its great similarity with WZ Sge.
V455 And showed grazing eclipses, therefore the orbital period is obtained as $P_{\rm orb} = 0.05630921(1)$ d, combining with the radial velocity observations.
The existence of grazing eclipses suggests that V455 And is highly inclined as much as $i \sim 75^\circ$. 
Moreover, time resolved studies have found that V455 And shows a spin period (67.3 s) of the primary WD, establishing that V455 And is an intermediate polar \citep{ara05v455and, sil12v455andGALEX, szk13V455And, gha20v455and}.
The first observed outburst of V455 And was discovered in 2007 by H. Maehara \footnote{<http://ooruri.kusastro.kyoto-u.ac.jp/mailarchive/vsnet-alert/9530>} and follow-up observations were conducted worldwide.
The optical photometric observations of this 2007 superoutburst and the evolution of the superhumps are summarized in \citet{Pdot, mat09v455and, mae09v455andproc}.
Using the period of early superhump and growing ordinary superhump (Stage-A superhump), the mass ratio of V455 And is dynamically estimated to be 0.080 \citep{kat13qfromstageA}.
In \citet{mat09v455and}, the multi-color photometry of the early superhump was performed for the first time, showing that the profile and color of early superhumps can be explained by the vertically extended outer disk.
Moreover, \citet{uem12ESHrecon} performed "early superhump modeling" using the multi-color light curve  of early superhumps of V455 And.
They successfully reconstructed the accretion disk with the two flaring patterns in the outermost region of the accretion disk, which are responsible for the maxima of early superhumps.
Along with these patterns, the flaring patterns are also elongated into the inner part of the accretion disk.
\citet{uem12ESHrecon} interpreted these structures in the accretion disk as a double-spiral structure emerging  from the 2:1 resonance.

In this paper, we present the spectroscopic observations of V455 And performed during the 2007 superoutburst and the subsequent post-superoutburst stage.
Section \ref{sec:2} overviews our observations, and Section \ref{sec:3} shows the results of our spectroscopic observations.
In Section \ref{sec:4}, we performed Doppler tomography of the data obtained during the early superhump phase.
We discuss the features and interpretations of our spectra in Section \ref{sec:5} and give a summary in Section \ref{sec:6}.

\section{Observations and Analysis}
\label{sec:2}

\begin{figure*}[tbp]
 \begin{center}
    \includegraphics[width=170mm]{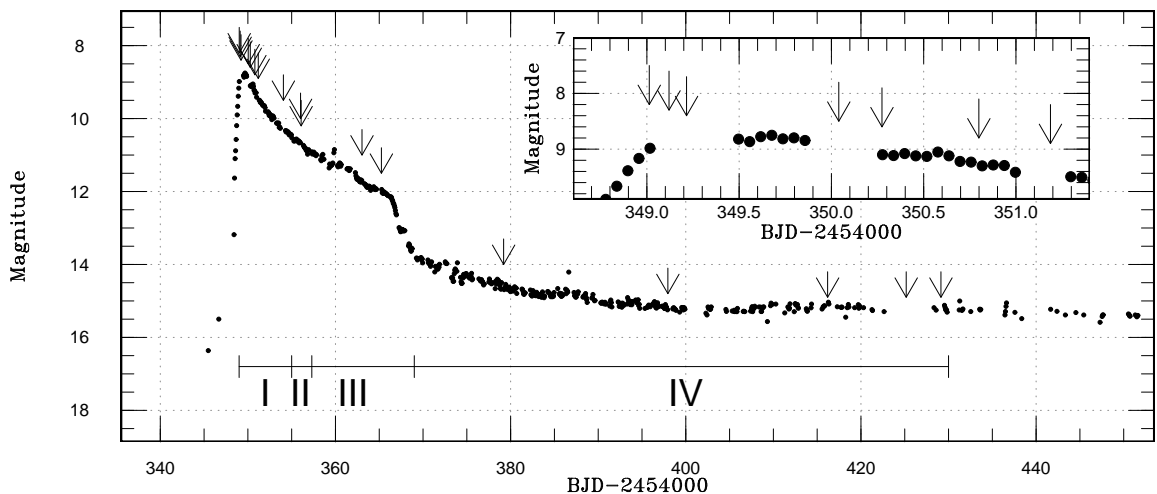}
  \end{center}
 \caption{
    Light curve of V455 And during its 2007 superoutburst, binned in each 0.06 d. 
    Data was taken from \citet{Pdot}.
    Arrows show the epoch of our spectroscopic observations.
    The upper right panel shows the enlarged light curve around the outburst peak.
    As shown at the bottom of this figure, we separate our observations into four periods (see Section \ref{sec:3}). 
    }
 \label{fig:oc-lc}
\end{figure*}

Our spectroscopic observations of V455 And were performed from around the optical peak to the post-superoutburst stage with various telescopes and equipment.
The log of our observations is summarized in Table \ref{tab:1}.
In Figure \ref{fig:oc-lc}, the light curve of the 2007 superoutburst of V455 And is presented with the black arrows indicating the epoch of our spectroscopic observations.
The light curve data was taken from \citet{Pdot}.
Most of our spectra were taken during the early superhump stage.
We obtained some spectra during the ordinary superhump phase and during the post-superoutburst stage as well (see Section \ref{sec:3} for details). 
The data reduction was performed using IRAF in the standard manner 
 (bias subtraction, flat fielding, aperture determination, spectral extraction, wavelength
calibration with arc lamps, flux calibration with standard stars, and normalization by the continuum).

\begin{longtable}{ccccc}
  \caption{Log of spectroscopic observations}\label{tab:1}
  \hline              
    Mid BJD & Exposure  & Number of  & Site & Spectral  \\ 
     &  Time & spectra &  & Range \\ 
     (2454000+) & (s) &  & & ( \AA)\\
    \hline
    \hline
\endfirsthead
\endfoot
349.01 & 600 &  7 & Fujii-Bisei    & 3800-8200 \\
349.12 & 300 & 15 & Bisei          & 3700-8200 \\
349.22 & 600 & 15 & Nishi-Harima   & 4470-4900 \\
350.04 & 300 &  3 & Bisei          & 3700-8200 \\
350.27 & 600 &  6 & Fujii-Bisei    & 3800-8200 \\
350.80 &  30 & 87 & Subaru         & 4200-5300\\ 
350.80 &  30 & 87 & Subaru         & 5600-6700\\ 
351.19 &  60 & 70 & Gunma          & 4000-8000 \\
354.05 & 600 &  9 & Fujii-Bisei    & 3800-8200 \\
356.01 & 720 &  7 & Fujii-Bisei    & 3800-8200 \\
356.08 & 600 &  4 & Bisei          & 3700-8200 \\
363.03 & 600 &  3 & Bisei          & 3700-8200 \\
365.25 & 120 & 60 & Gunma          & 4000-8000 \\
379.20 & 150 & 45 & Gunma          & 4000-8000 \\
397.96 & 300 &  3 & Gunma          & 4000-8000 \\
416.19 & 180 & 15 & Gunma          & 4000-8000 \\
425.17 & 180 & 22 & Gunma          & 4000-8000 \\
429.13 & 180 & 35 & Gunma          & 4000-8000 \\
\hline
\end{longtable}

\subsection{Bisei Astronomical Observatory}

We took the low resolution spectra (R $\sim$1,400) with the spectrograph mounted on the 101-cm telescope at the Bisei Astronomical Observatory (BAO) on BJD 2454349.12, 2454350.04, 2454356.08, and 2454363.03.
The exposure times were 300 s for the first two nights and 600 s for the last two nights, respectively.
%http://www.bao.city.ibara.okayama.jp/eng/kikigaiyou.html

\subsection{Subaru telescope and the High Dispersion Spectrograph}
On BJD 2454350.80, the high-resolution spectra (R $\sim 40,000$) were obtained with the High Dispersion Spectrograph (HDS; \cite{nog02HDS, HDS2}) mounted on the 8.2-m Subaru telescope at the Maunakea Observatory.
The exposure time was 30 s.
We have obtained 87 exposures with the wavelength coverage of 4200-5300 \AA~ and 5600-6700 \AA, simultaneously covering both He II 4686 \AA~ and H$\alpha$. 
%https://www.subarutelescope.org/Observing/Instruments/HDS/index.html

\subsection{Fujii-Bisei Observatory}

The low resolution spectra (R$\sim$ 600) were obtained on BJD 2454349.01, 2454350.27, 2454354.05, and 2454356.01, with the spectrograph, FBSPEC-2, mounted on a 28-cm Schmidt-Cassegrain telescope at the Fujii-Bisei Observatory.
The exposure times were 600 s on the first three observation nights, and 720 s on the last night, respectively.

\subsection{Gunma Astronomical Observatory}

The low resolution spectra (R $\sim$ 500)  were taken with the Gunma LOW resolution Spectrograph (GLOWS) mounted on a 150-cm Ritchey-Chretien telescope at the Gunma Astronomical Observatory (GAO). 
Our observations were performed on BJD 2454351.19, 2454365.25, 2454379.20, 2454397.96, 2454416.19, 2454425.17, and 2454429.13, and the exposure times were 60, 120, 150, 300, 180, 180, 180 s, respectively.
On the first two nights V455 And was in the superoutburst, and on the other nights it was in the post-superoutburst stage.
%http://www.astron.pref.gunma.jp/e/telescopes.html

\subsection{Nishi-Harima Astronomical Observatory}
We obtained the mid-resolution spectra (R $\sim 7,500$) on BJD 2454349.22, using the Medium And Low-dispersion Long-slit Spectrograph (MALLS; \cite{NayutaMALLS}) mounted on the 2.0-m NAYUTA telescope of the Nishi-Harima Astronomical Observatory.
The integration time was 600 s, and the wavelength coverage is 4470-4900 \AA.
%http://www.nhao.jp/research/instruments.html
%http://www.nhao.jp/~malls/malls_wiki/index.php

\section{Result}
\label{sec:3}

\begin{figure*}[tbp]
 \begin{center}
    \includegraphics[width=165mm]{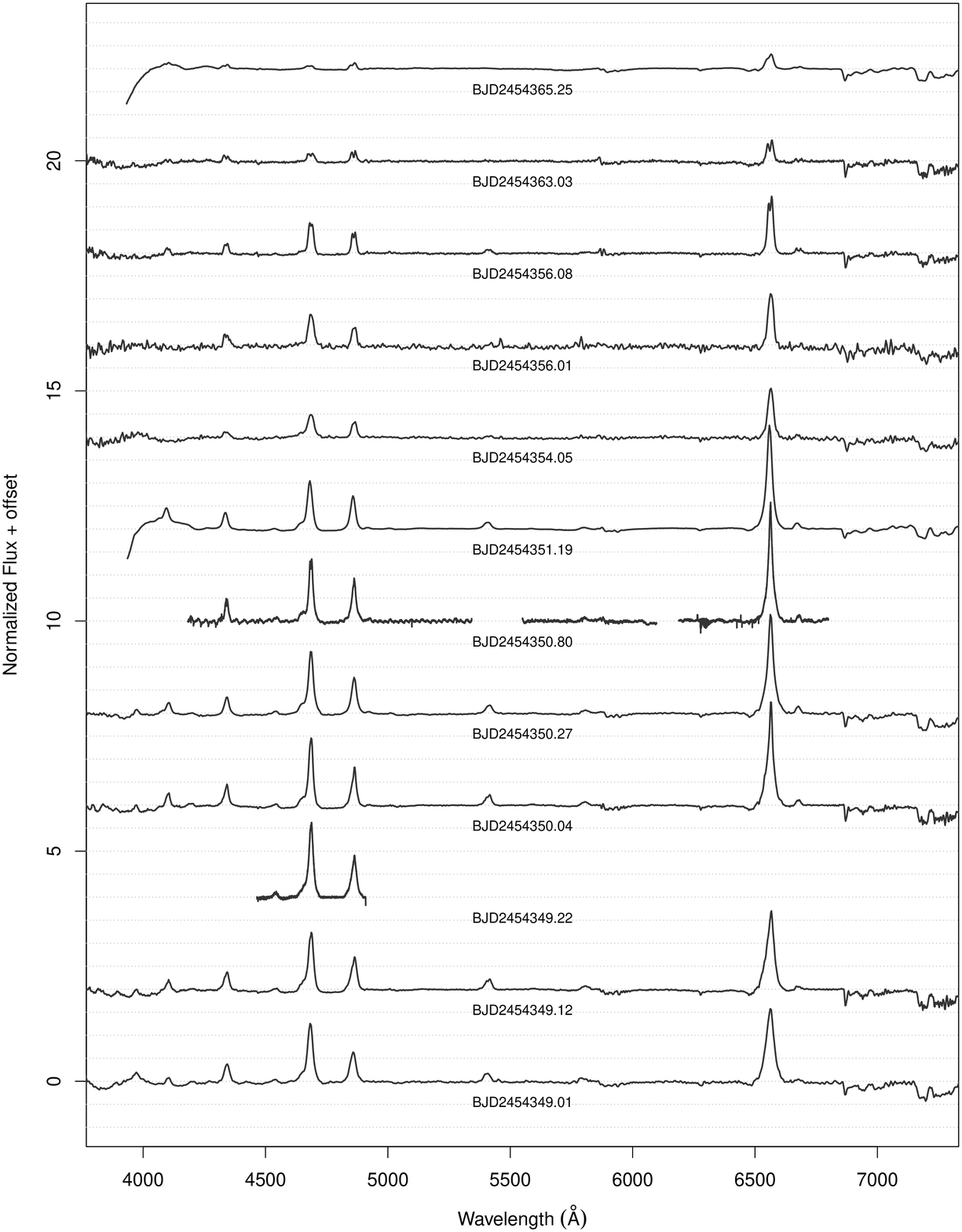}
  \end{center}
 \caption{
 Normalized spectra taken during the superoutburst.
 For visibility, the data are vertically shifted.}
 \label{fig:spec-outburst}
\end{figure*}

\begin{figure}[tbp]
 \begin{center}
    \includegraphics[width=83mm]{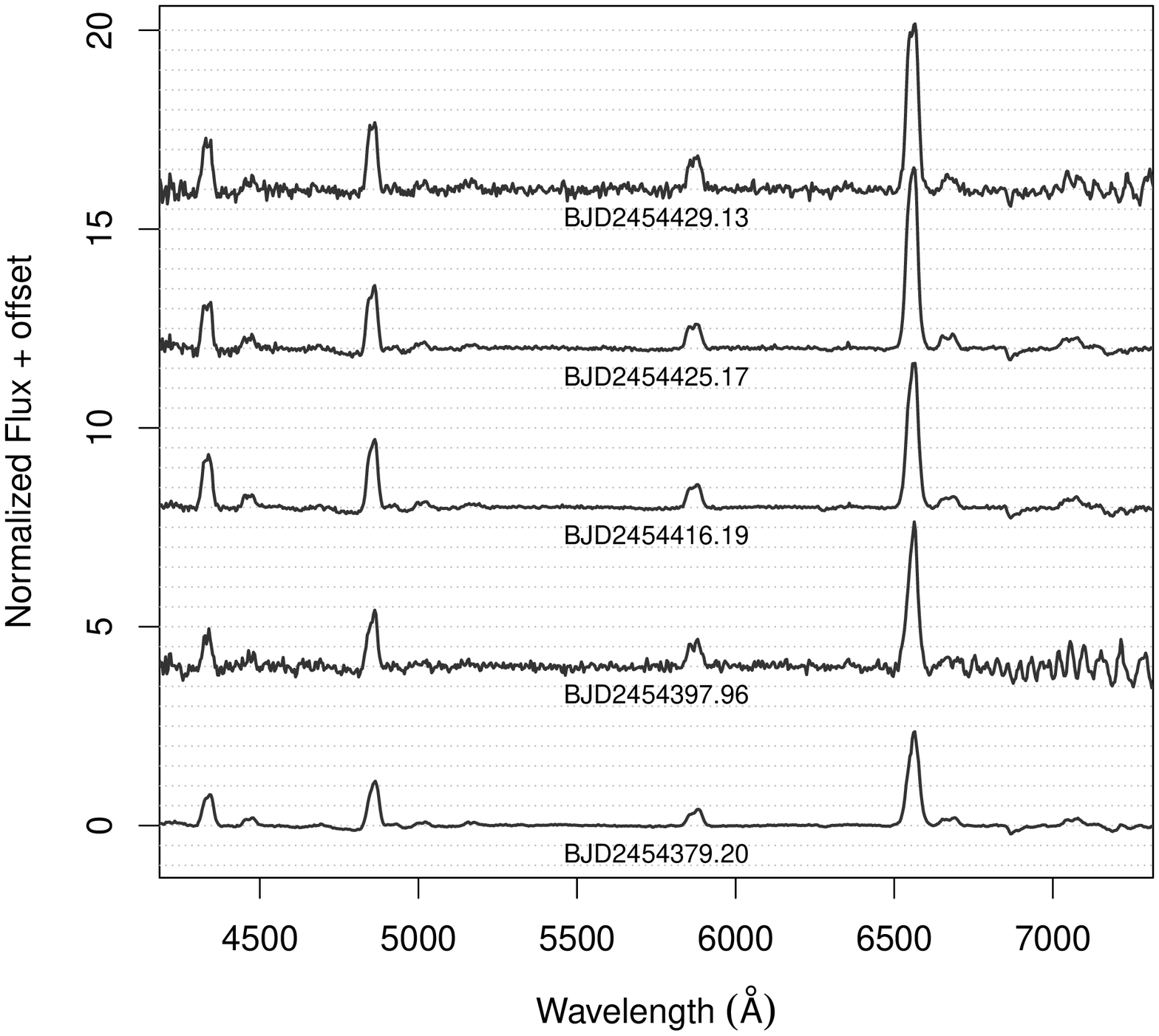}
  \end{center}
 \caption{
 Normalized spectra taken during the post-superoutburst stage.
  For visibility, the data are vertically shifted.}
 \label{fig:spec-post}
\end{figure}

In this section, we present the characteristics and evolution of our spectra of V455 And taken during the 2007 superoutburst and the following post-superoutburst stage.
In Figure \ref{fig:spec-outburst}, we show all our normalized spectra taken during the superoutburst, and Figure \ref{fig:spec-post} presents the spectra taken during the post-superoutburst stage.
The equivalent widths (EWs) of our spectra are listed in Table \ref{tab:2} (Balmer series, He II, C IV / N IV blend) and Table \ref{tab:3} (He I series).
Note that since  EW is defined as a positive value for an absorption line, the negative value is given for an emission line.

In our analyses, our observations are divided into four periods based on both spectroscopic and photometric behaviour \citep{Pdot, mat09v455and}.
Period I corresponds to the early superhump phase, which is BJD 2454349.0 - 2454355.0.
During this Period I, single-peaked Balmer emission lines were observed.
Period II is BJD 2454355.0 - 2454357.3, during which the superposition of early and ordinary superhumps was observed.
The spectra in Period II showed Balmer series with double-peaked emission, and the emission of He II 4686 \AA~ was still stronger than H$\beta$.
Period III covers the ordinary superhump phase, BJD 2454357.3 - 2454369.
In Period III, the high excitation lines such as He II and C IV / N IV blend were significantly weaker than Period I and II.
Period IV is during the post-superoutburst stage corresponding the observations after BJD 2454369.

\begin{longtable}{r|rrrrrrr}
  \caption{Averaged equivalent widths in  \AA~ of absorption lines (negative EW for an emission line).}\label{tab:2}
  \hline              
    Mid BJD     & H$\alpha$ & H$\beta$ & H$\gamma$  & He II   & He II   & He II   & C IV  \\ 
    (2454000+)  &           &          &            & 4542 \AA~ & 4686 \AA~\commenta & 5411 \AA~ & N IV \\ 
    \hline
    \hline
\endfirsthead
    \multicolumn{8}{l}{\commenta{} Can be contaminated with Bowen blend.}\\
\endfoot
349.01 & -65.22 (1.5) & -19.08 (0.52) & -8.55 (0.52) & 
-4.60 (0.46) & -39.05 (0.70) & -5.99 (0.22) & -4.55 (0.35) \\
349.12 & -64.13 (1.24) & -19.17 (0.37) & -10.90 (0.39) &
-1.88 (0.10) & -32.41 (0.34) & -7.68 (0.16) & -3.57 (0.23)\\
349.22 &  & -24.40 (0.56) &  & -1.94 (0.28) & -43.63 (0.49) &  &  \\
350.04 & -68.05 (1.43) & -25.15 (0.63) & -11.57 (0.47) 
& -2.66 (0.23) & -26.96 (0.33) & -6.15 (0.13) & -3.47 (0.22) \\
350.27 & -73.31 (0.60) & -23.75 (0.52) & -11.32 (0.33) & 
-2.70 (0.31) & -39.54 (0.56) & -7.16 (0.31) & -4.47 (0.18) \\
350.80 & -63.36 (1.06) & -22.60 (0.81) & -6.09 (0.40) & -1.24 (0.15) & -32.72 (0.65) &  & -1.27 (0.10) \\
351.19 & -72.39 (0.58) & -21.21 (0.19) & -8.56 (0.47) &
-1.29 (0.11) & -24.62 (0.11) & -4.83 (0.08) & -2.44 (0.29) \\
354.05 & -34.74 (0.72) & -9.09 (0.24) & -3.15 (0.51) &
-0.88 (0.31) & -15.58 (0.38) &  &  \\
356.01 & -40.75 (0.6) & -10.16 (0.42) & -8.04 (0.22) &  & -17.56 (0.96) &  & \\
356.08 & -35.92 (0.5) & -11.46 (0.33) & -5.56 (0.11) &
-0.62 (0.05) & -15.43 (0.14) & -2.51 (0.07) & -1.06 (0.15) \\
363.03 & -14.36 (0.60) & -6.28 (0.29) & -3.59 (0.27) & & -5.28 (0.20) &  &  \\
365.25 & -12.27 (0.36) & -3.99 (0.13) & -1.50 (0.13) &  & -2.80 (0.10) &  &  \\
\hline
379.20 & -90.98 (0.59) & -49.91 (0.89) & -27.78 (0.63) &  &  &   &  \\
397.96 & -131.3 (2.26) & -52.48 (1.8) & -18.81 (1.58) &  &  &   &  \\
416.19 & -137.21 (1.07) & -84.01 (1.34) & -46.97 (1.54) &  &  &   &  \\
425.17 & -174.23 (1.07) & -77.28 (1.68) & -39.44 (1.01) &  &  &   &  \\
429.13 & -176.2 (2.71) & -79.62 (2.37) & -57.33 (1.92) &  &  &   &  \\
\hline
\end{longtable}

\begin{longtable}{r|rrrrrr}
  \caption{Averaged equivalent widths in  \AA~ of absorption lines (negative EW for an emission line).}\label{tab:3}
  \hline              
    Mid BJD     & He I & He I  & He I   & He I   & He I   & He I  \\ 
    (2454000+)  & 4471 \AA~  & 4922 \AA~  & 5015 \AA~ & 5876 \AA~ & 6678 \AA~ & 7065 \AA~  \\ 
    \hline
    \hline
\endfirsthead
\endfoot
349.01 & 1.09 (0.21) & ---- & ---- & 4.77 (0.23) & -0.43 (0.08) & ----  \\
349.12 & 0.51 (0.11) & ---- & ---- & 1.24 (0.14) & -1.86 (0.09) & ----  \\
349.22 & ---- & ---- & ---- & ---- &---- & ----  \\
350.04 & 0.08 (0.02) & -0.71 (0.04) & -0.19 (0.05) & ---- & -2.95 (0.12) & ----  \\
350.27 & ---- & -0.33 (0.17) & -0.30 (0.06) & -0.66 (0.06) & -2.86 (0.08) & -1.11 (0.1)  \\
350.80 & -1.07 (0.08) & -0.67 (0.07) & ---- & -0.51 (0.09) & -1.82 (0.09) & ----  \\
351.19 & 0.07 (0.01) & -0.52 (0.03) & -0.32 (0.02) & -0.31 (0.02) & -3.25 (0.11) & -1.28 (0.2)  \\
354.05 & ---- & ---- & ---- & ---- & -1.90 (0.13) & ----  \\
356.01 & ---- & ---- & ---- & -1.05 (0.27) &---- & ----  \\
356.08 & -0.4 (0.06) & ---- & ---- & -0.53 (0.08) & -3.33 (0.16) & -1.46 (0.07)  \\
363.03 & 0.59 (0.07) & 0.46 (0.06) & 0.19 (0.02) & ---- & -0.99 (0.1) & ----  \\
365.25 & ---- & ---- & ---- & ---- & -1.82 (0.04) & ----  \\
\hline
379.20 & -8.50 (0.40) & ---- &  -6.28 (0.24) &
-19.59 (0.26) & -10.93 (0.71) & -9.88 (0.59)  \\
397.96 & ---- & ---- & ---- & -31.74 (1.40) &---- & ----  \\
416.19 & -9.85 (0.42) & ---- & -8.88 (0.40) & -24.61 (0.64) & -12.82 (0.97) & -15.13 (0.80)  \\
425.17 & -19.97 (1.02) & -2.10 (0.89) & -5.04 (0.31) & -31.09 (0.87) & -16.63 (0.65) & -14.82 (0.87)  \\
429.13 & -17.63 (1.52) & ---- & -12.31 (0.88) & -28.64 (1.55) & -27.25 (2.39) & ----  \\
\hline
\end{longtable}

\subsection{Period I: early superhump phase}

\begin{figure*}[tbp]
 \begin{center}
    \includegraphics[width=170mm]{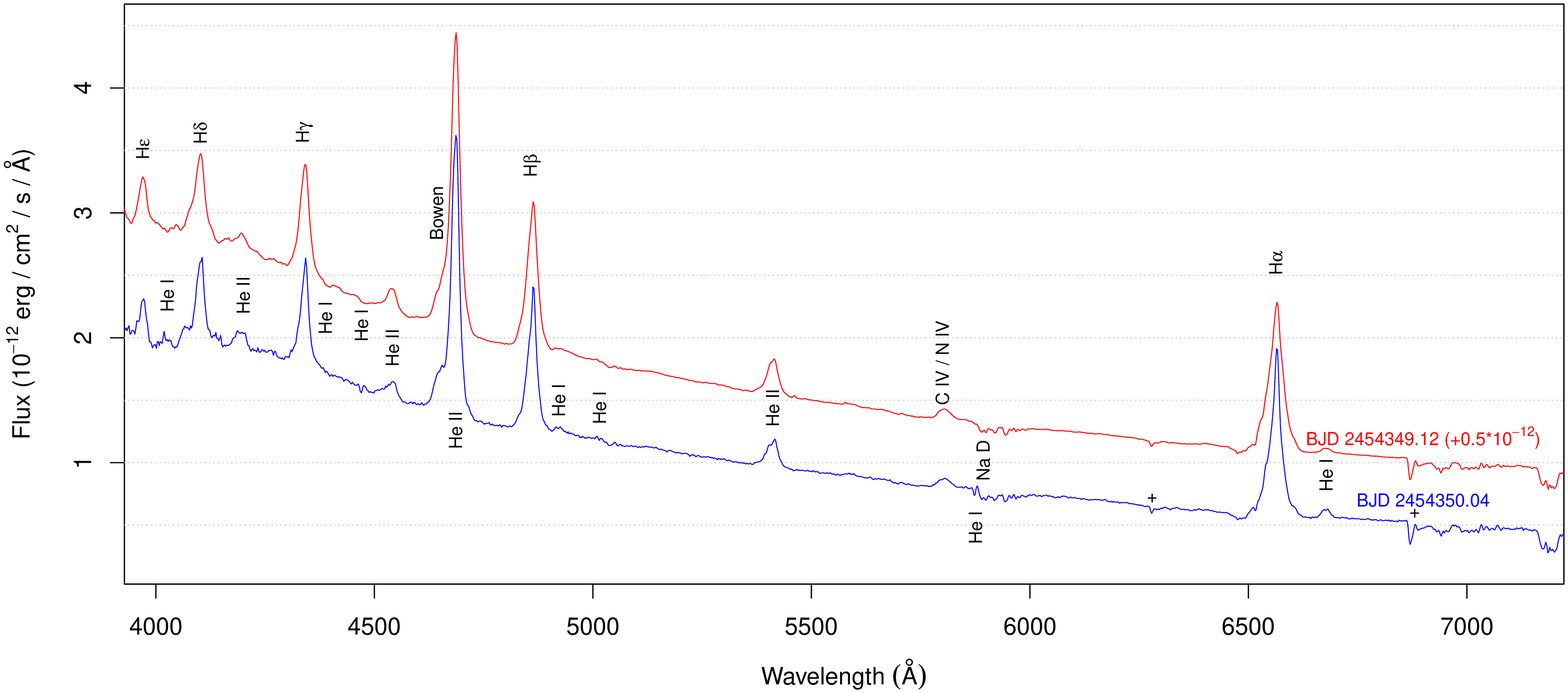}
  \end{center}
 \caption{
 Period I spectra taken on BJD 2454349.12 (upper red, vertically shifted) and BJD 2454350.04 (lower blue)  at BAO.}
 \label{fig:period.1}
\end{figure*}

\begin{figure*}[tbp]
 \begin{center}
    \includegraphics[width=83mm]{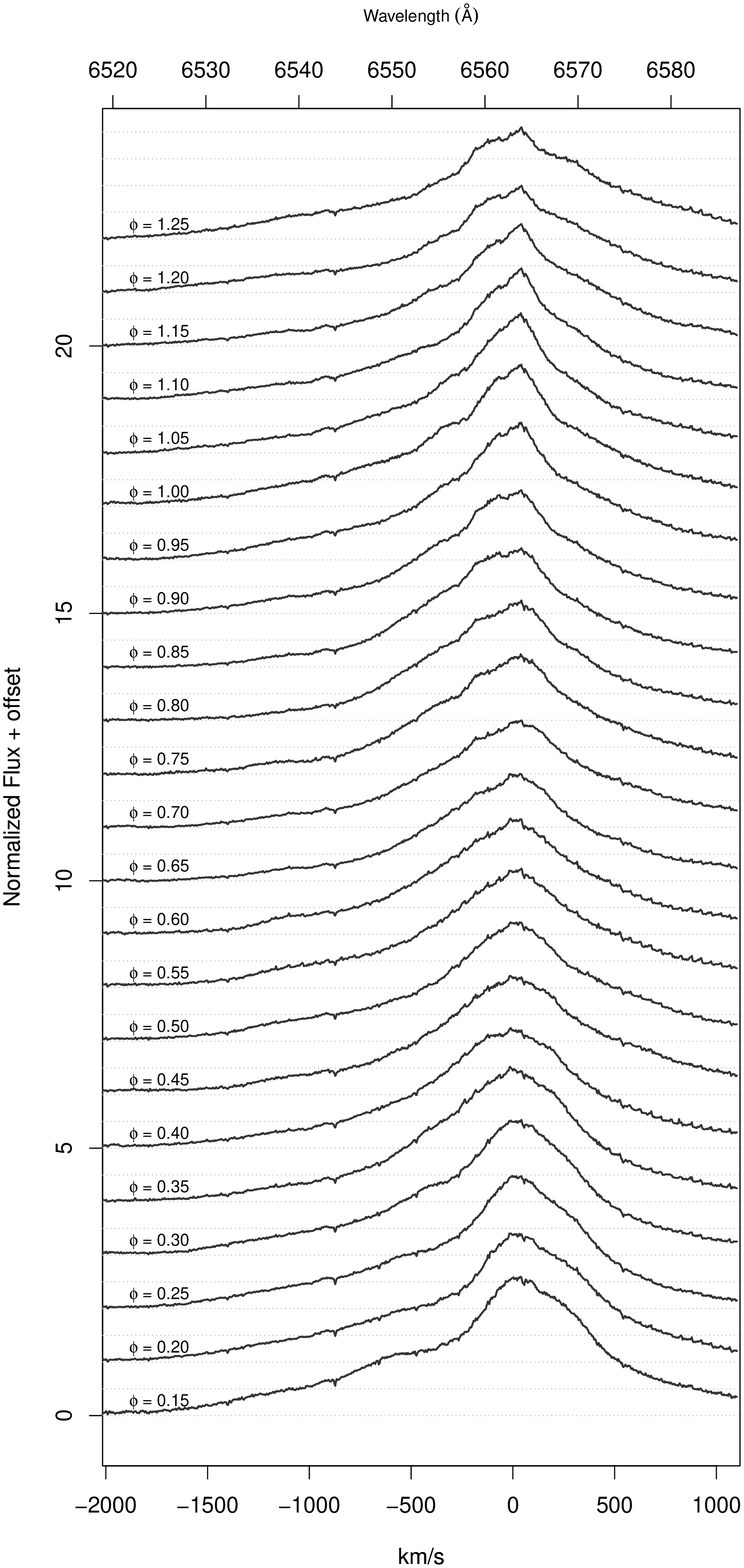}
    \includegraphics[width=83mm]{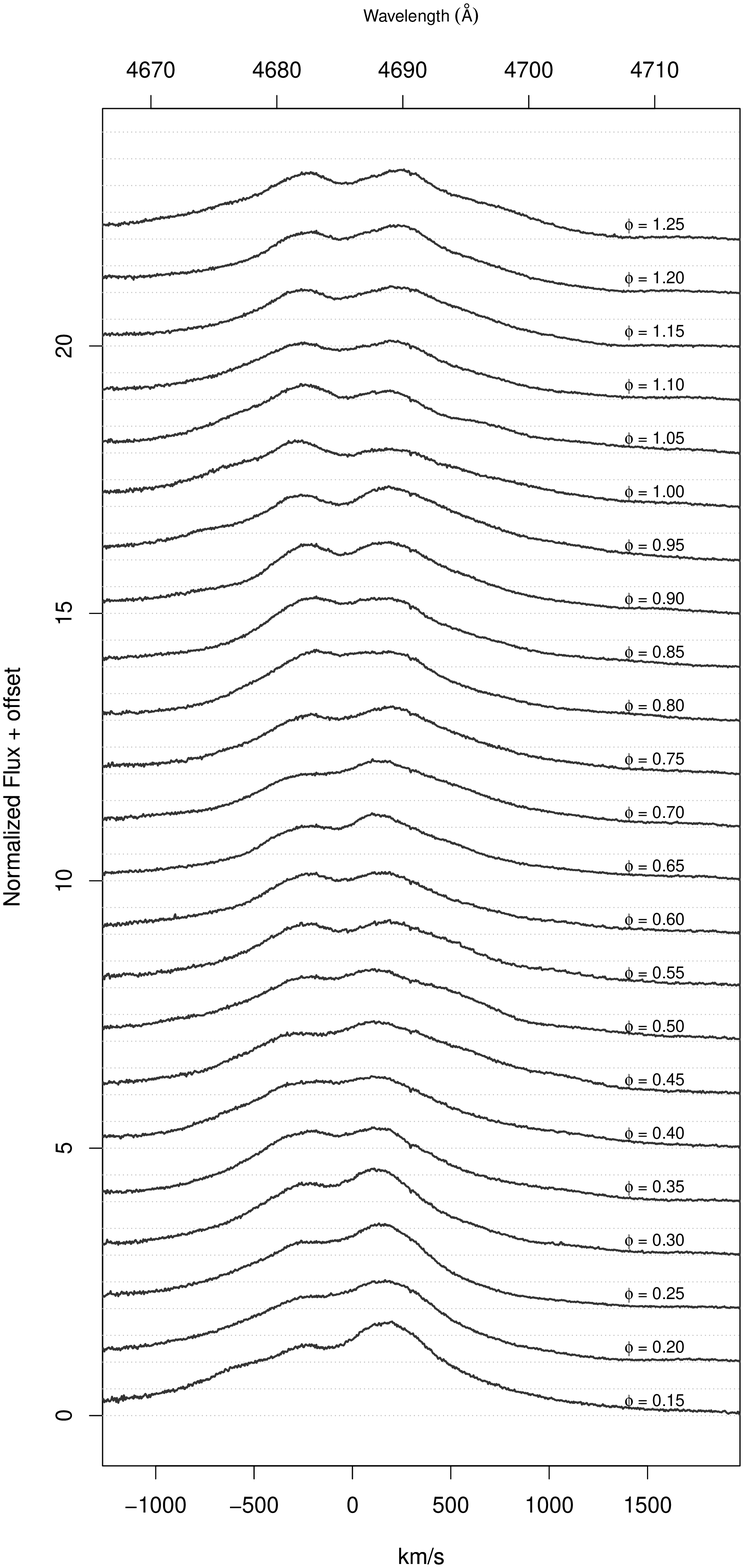}
  \end{center}
 \caption{
 Phase-averaged spectra of H$\alpha$ (left panel) and He II 4686 \AA~ (right panel) taken with HDS mounted on Subaru Telescope on BJD 2454350.80.
 $\phi = 1.0$ is defined as BJD 2454350.81583 \citep{ara05v455and}. }
 \label{fig:phase-lines}
\end{figure*}

As described above, the spectra in Period I were observed during the early superhump phase.
In Figure \ref{fig:period.1}, the  spectra on BJD 2454349.12 (upper red) and 2454350.04 (lower blue) obtained at BAO.
Figure \ref{fig:phase-lines} presents the phase-resolved emission lines of H$\alpha$ (left) and He II 4686 \AA~(right) on BJD 2454350.80 observed with  HDS.
In our analyses, 0.01 phase was added to the ephemeris of \citet{ara05v455and}, following the suggestion in \citet{uem12ESHrecon}.
These spectra are binned in each 0.05 phase in Figure \ref{fig:phase-lines}, and the phase 1.0 is corresponding to BJD 2454350.81583.
Strong Balmer and He II 4686 \AA~ emission lines are recognized in our spectra.
Along with these lines, He I, Pickering series of He II 4200 \AA, 4542 \AA, 5411 \AA,  C$_{\rm III}$ / N$_{\rm III}$ Bowen blend, and  C$_{\rm IV}$ / N$_{\rm IV}$ blend emission lines were detected.
Although V455 And is a high-inclination system \citep{ara05v455and}, Balmer series showed a single-peaked profile .
The full width at zero intensity (FWZI) of H$\alpha$ is about {4,000} km/s, which is comparable with the typical value in eclipsing CVs \citep{war95book}.%
As seen in Figure \ref{fig:phase-lines}, both in H$\alpha$ and He II 4686 \AA, the line profiles were almost stable around and in the eclipse ($\phi = 1$), indicating the emission regions of both lines are vertically extended above the disk plane.
On  BJD 2454349, He I 6678 line is in double-peaked  emission with the peak separation of $\sim 430$ km/s, while the other He I series are observed as absorption lines or not detected.
One day after, on BJD 2454350, He I 4922 \AA~ and 5015 \AA~ are likely turned into  emission.
He II 5412 \AA~ as well shows a double-peaked emission line, with the peak separation of $\sim 700$ km/s.
In the right panel of Figure \ref{fig:phase-lines}, He II 4686 \AA~ emission line shows a double peak emission with the peak separation of $\sim {440}$ km/s, which is much narrower than He II 5412 \AA.
The asymmetric bump on the blue side of He II 4686 \AA~ emission line implies that Bowen blend line was also in the emission.
C$_{\rm IV}$ / N$_{\rm IV}$ blend and He II 5412 \AA~ were also detected during the early superhump phase of WZ Sge \citep{nog04wzsgespec}.
On the other hand, He II 4200 and 4542 line are rarely detected in DN outbursts (e.g., HT Cas in outburst; \cite{neu20HTCasoutburstdopmap}), though often observed in intermediate polars \citep{sch91amher, har99J0558}.
Moreover, in the left panel of Figure \ref{fig:phase-lines}, a small bump centered around 6560 \AA~ is seen in H$\alpha$ emission line.
This bump is likely an emission line from Pickering series 6560 \AA.
Therefore, the other Balmer series can be contaminated with Pickering series as well.

\subsection{Period II: early superhump - ordinary superhump transition}

\begin{figure}[tbp]
 \begin{center}
    \includegraphics[width=83mm]{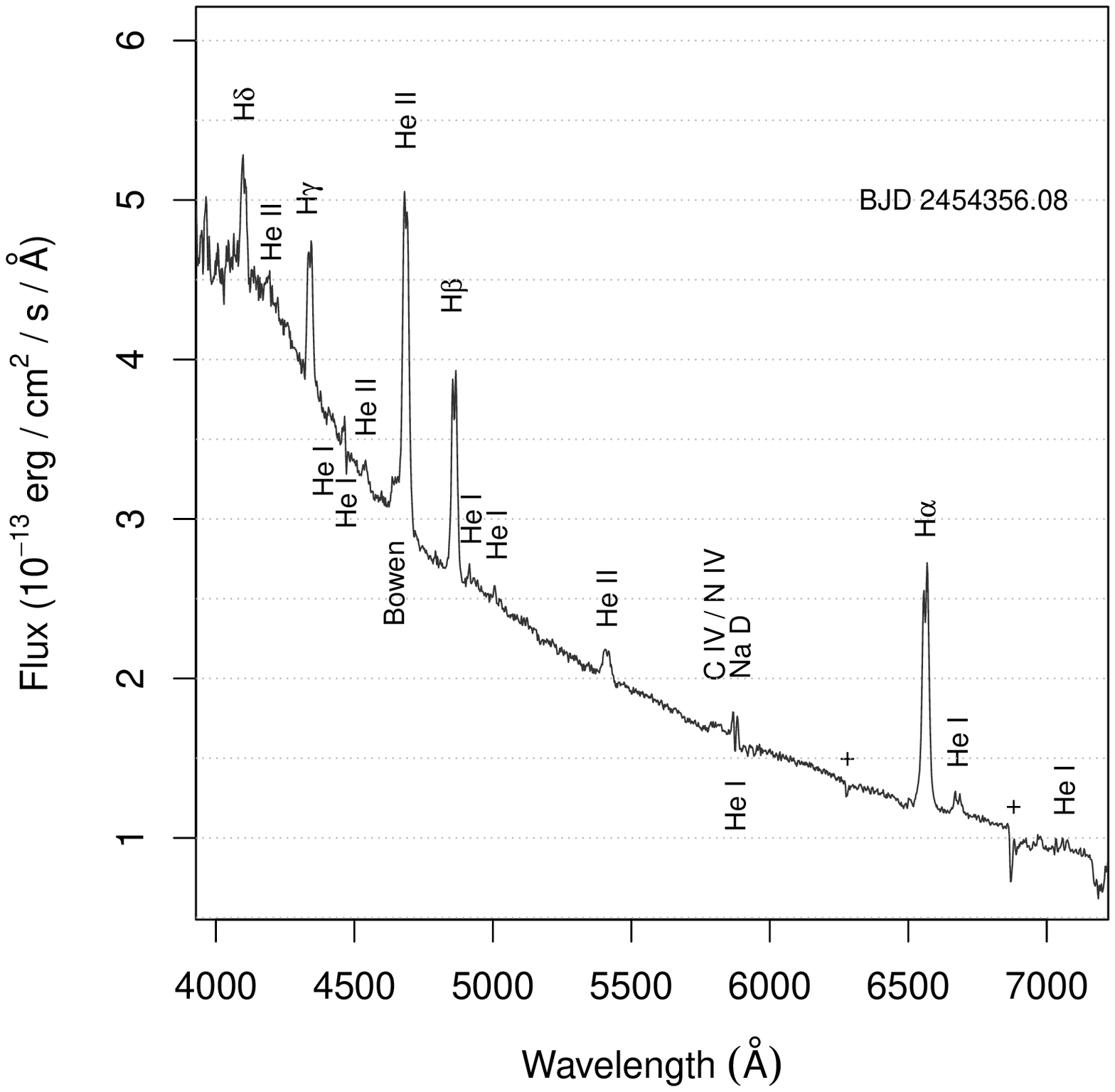}
  \end{center}
 \caption{
 Period II spectrum taken on BJD 2454356.08 at BAO.}
 \label{fig:period.2}
\end{figure}

Period II is a transition phase from early superhumps to ordinary superhumps.
The  spectrum taken on BJD 2454356.08 at BAO is presented in Figure \ref{fig:period.2}.
Compared to Period I, the emission lines were significantly weakened and Balmer series turned into double-peaked  emission.
The peak separations of H$\alpha$,  H$\beta$, and H$\gamma$ are $\sim 530$ km/s, $\sim 660$ km/s, and $\sim 600$ km/s, respectively.
He I lines are in emission, in contrast to Period I. 
He I 6678 \AA~ is obviously in double-peak, with the peak separation of $\sim 750$ km/s. 
He II Pickering series, Bowen blend and C$_{\rm IV}$ / N$_{\rm IV}$ emission lines are detected, and He II 4686 \AA~ emission line is still stronger than H$\beta$.
The peak separations of He II 4686 \AA~ and 5412 \AA~ are $\sim 610$ km/s and $\sim 660$s km/s, respectively.

\subsection{Period III: ordinary superhump phase}

\begin{figure}[tbp]
 \begin{center}
    \includegraphics[width=83mm]{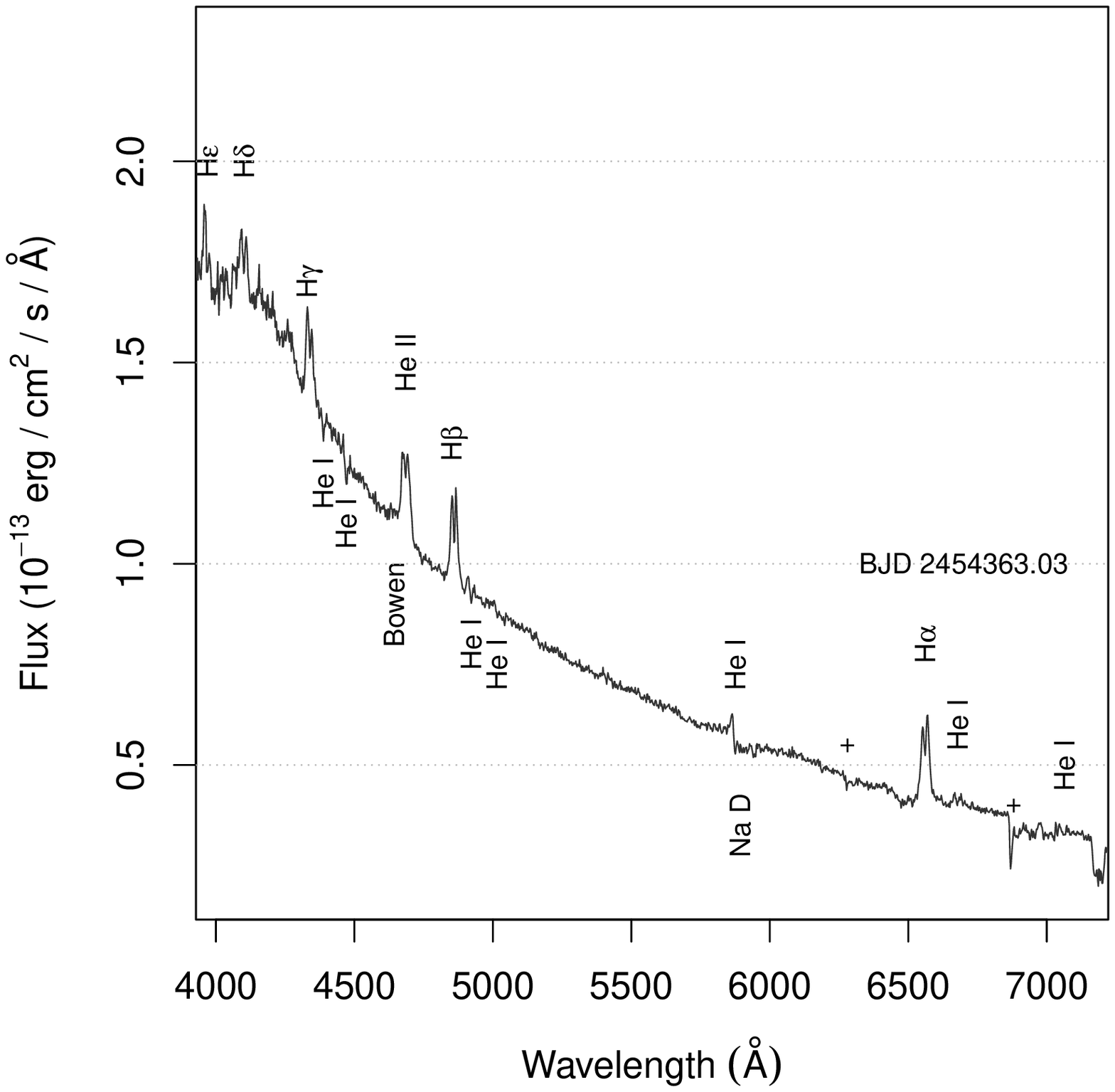}
  \end{center}
 \caption{
 Period III spectrum taken on BJD 2454363.03 at BAO.}
 \label{fig:period.3}
\end{figure}

This Period III includes two spectra taken during the ordinary superhump phase.
The  spectrum taken on BJD 2454363.03 at BAO is presented in Figure \ref{fig:period.3}.
The spectra in this period show double-peaked  emission lines from Balmer series and He II 4686 \AA~ with much wider peak separations than Period II.
The peak separations of H$\alpha$, H$\beta$, H$\gamma$, and He II 4686 \AA~ were $\sim 740$ km/s, $\sim 810$ km/s, $\sim 990$ km/s and $\sim 1,200$ km/s, respectively.
The EW of He II 4686 \AA~ emission line was smaller than that of H$\beta$ in this Period III.
He I lines are likely in absorption except He I 5876 \AA~ and 6678 \AA~ in weak emission.
Bowen blend is weakened from Period I and II, however still detected.
C$_{\rm IV}$ / N$_{\rm IV}$ blend nor He II 4200, 4542, 5411 lines were not detected in Period III.

\subsection{Period IV: post-superoutburst stage}

\begin{figure}[tbp]
 \begin{center}
    \includegraphics[width=83mm]{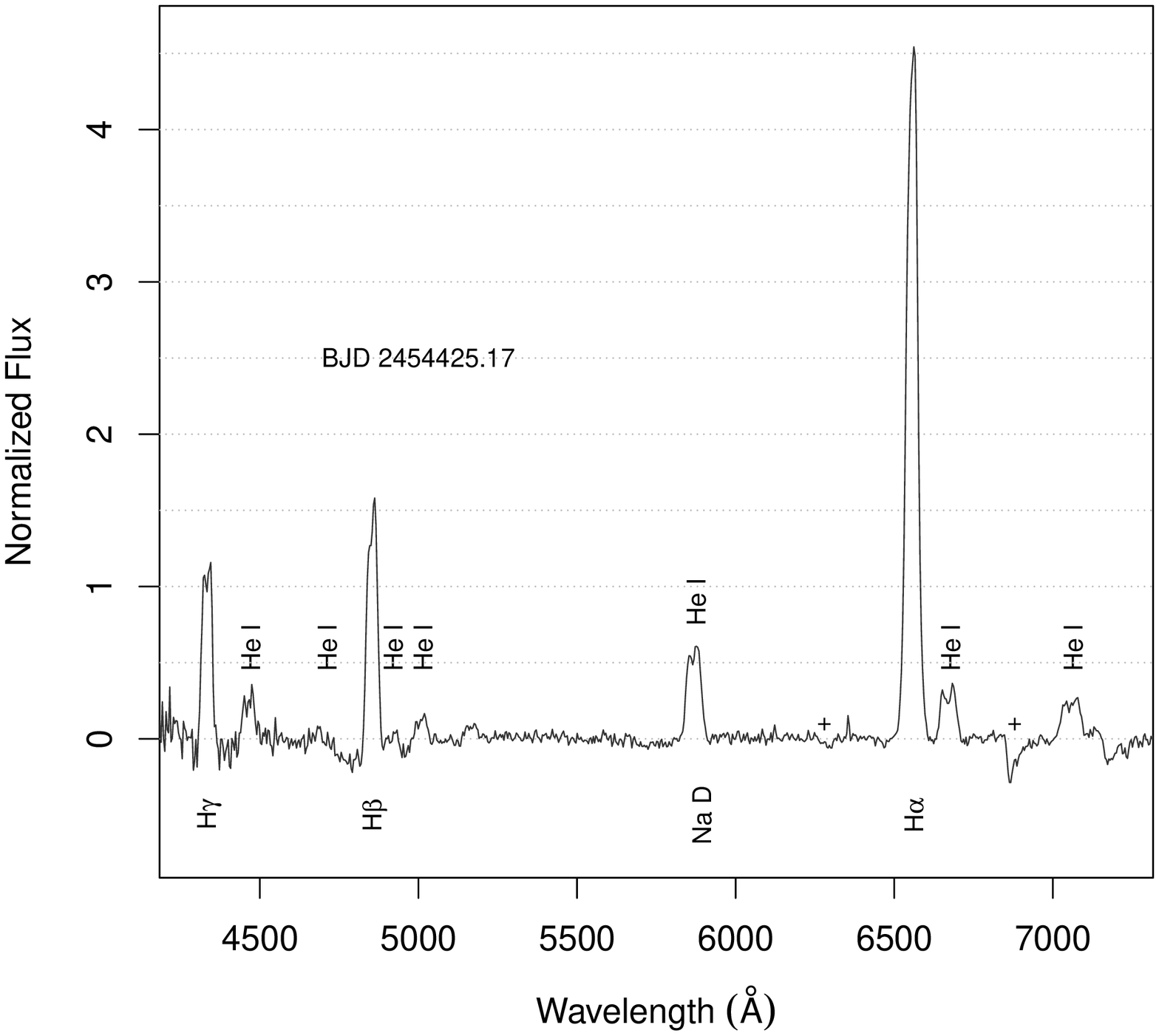}
  \end{center}
 \caption{
 Period IV spectrum taken on BJD 2454425.17 at GAO.}
 \label{fig:period.4}
\end{figure}

All spectra belonging to Period IV were obtained during the post-superoutburst stage at GAO.
The corresponding optical magnitude is around 15.5 mag, which is still $\sim$ 0.7 mag brighter than the quiescence brightness before the superoutburst \citep{ara05v455and}.
The normalized spectrum taken on BJD 2454425.17 is presented in Figure \ref{fig:period.4}.
The spectra during Period IV show double-peaked emission lines of Balmer series and He I.
The mean of peak separations of H$\beta$, H$\gamma$, and He I 6678  \AA~ are $\sim 950$ km/s, $\sim 1,085$ km/s, and $\sim 1,140$ km/s, respectively.
These values are broader than Period III, consistent with the spectrum of quiescence.
He II, Bowen blend nor C$_{\rm IV}$ / N$_{\rm IV}$ lines, observed during the superoutburst, were not detected in Period IV.

\section{Doppler Tomography}
\label{sec:4}

\begin{figure}[tbp]
 \begin{center}
    \includegraphics[width=83mm]{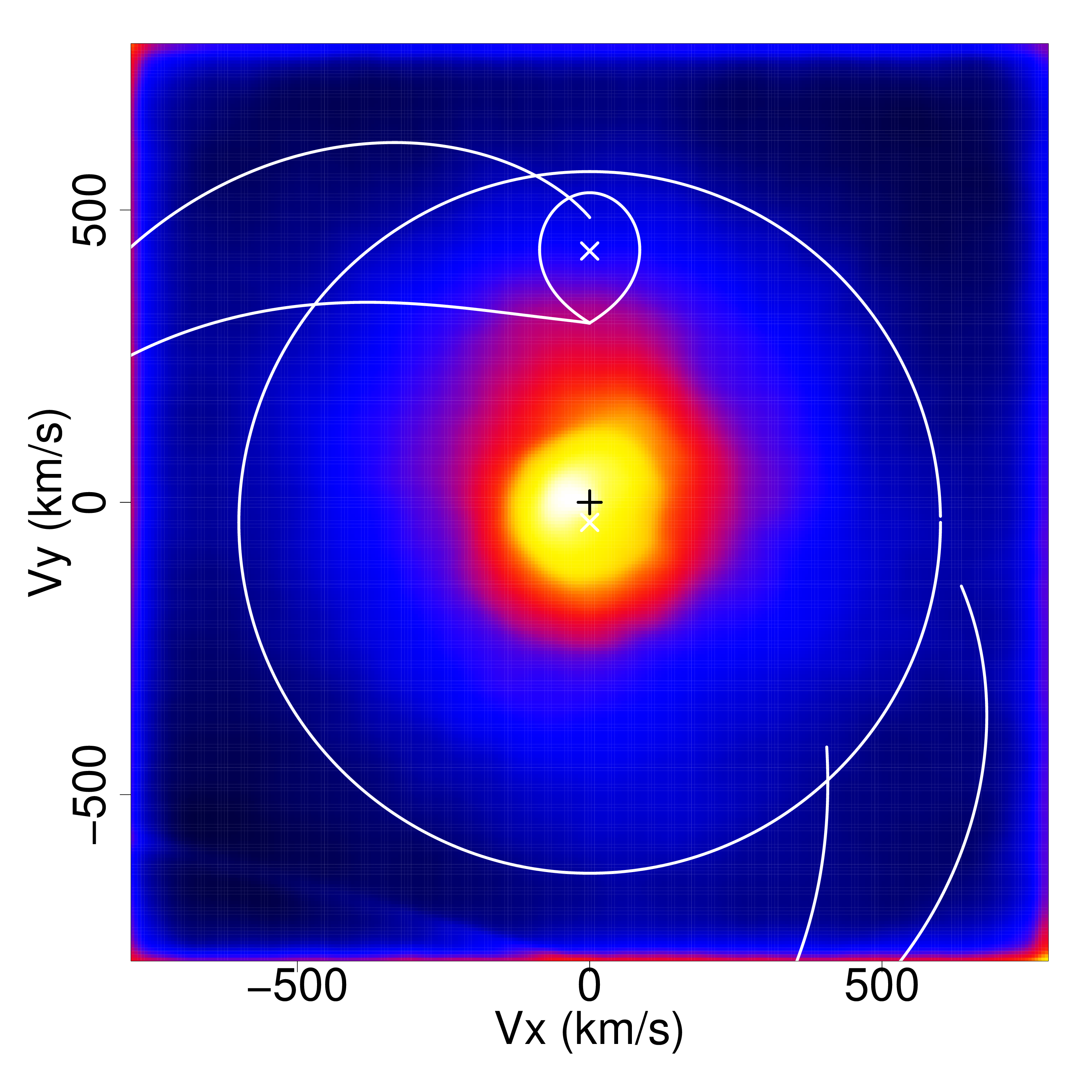}
    \includegraphics[width=83mm]{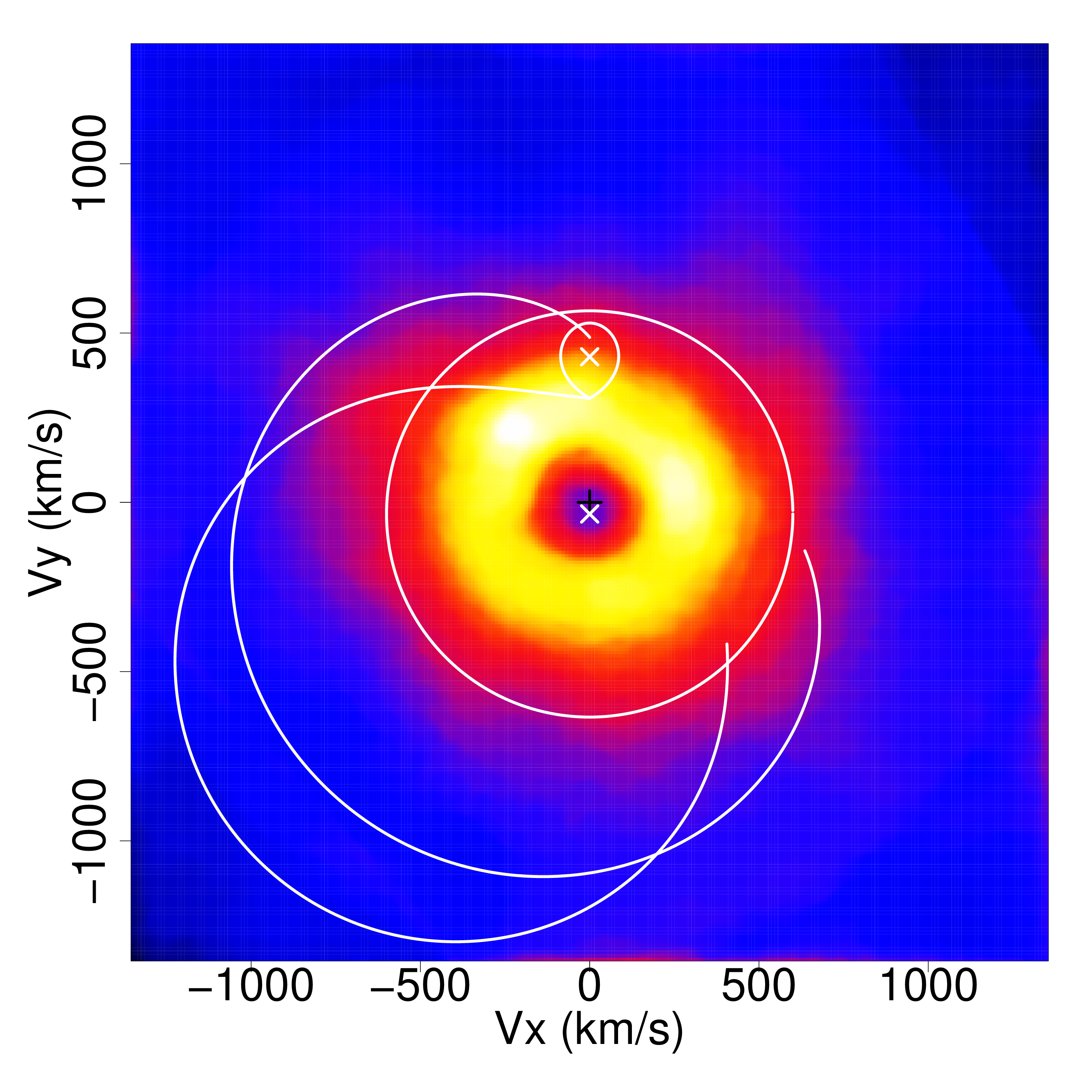}
    \end{center}
 \caption{
 Doppler tomography of H$\alpha$ (upper) and He II 4686 \AA~ (lower) on BJD 2454350.
 Secondary Roche lobe, Tidal truncation radius, and mass stream trajectory are plotted.
 Black and white crosses represent the center of mass of the binary and positions of the WD and secondary.
}
 \label{fig:dopmap}
\end{figure}

Doppler tomography is a method to reconstruct a brightness map of a certain line in the velocity plane firstly developed by \citet{DopplerTomography}.
In this paper, we used a code developed in \citet{uem15DTTVM}\footnote{<https://home.hiroshima-u.ac.jp/uemuram/dttvm/>}, which applies the total variation minimization technique to reconstruct the Doppler map.
For the system parameters to perform Doppler tomography, we used the following value from \citet{ara05v455and, kat13qfromstageA}; orbital period $P_{\rm orb} = 0.0.05630921$ d, inclination $i = 75^\circ$, mass ratio $q = 0.80$, primary WD mass $M_{\rm WD} = 0.6 M_{\odot}$.
As described in Section \ref{sec:3}, 0.01 phase was added to the orbital ephemeris of \citet{ara05v455and} as suggested by \citet{uem12ESHrecon}.
We note that, the data before/after 0.05 phase of the mid-eclipse are not used in performing Doppler Tomography following the usual manner.
In Figure \ref{fig:dopmap}, we show the Doppler map of H$\alpha$ (upper) and He II 4686 \AA~ (lower) using the data on  BJD 2454350 obtained with HDS.
In the Doppler map we show the secondary Roche lobe (inner distorted circle), and a mass transfer trajectory using the same system parameters.
The corresponding velocity of the tidal truncation radius ($R_{\rm max} \sim 0.6a$ where $a$ is a binary separation; \cite{pac77ADmodel}) under the assumption of Kepler rotation around the primary WD is plotted as the outer ring ($\sim$600 km/s).

In  the Doppler map of both H$\alpha$ and He II 4686 \AA, we do not see any structure likely related to a hot spot along the mass stream trajectory.
Therefore, a mass transfer burst was not taking place at this stage of the superoutburst.
The Doppler tomography of H$\alpha$ exhibits a rather compact blob centered at the primary WD.
The blob in the Doppler tomography of H$\alpha$ does not likely originate from an accretion disk, which would create a ring-like structure outside the tidal truncation radius in the Doppler map. 
This view is consistent with the spectra presented in the left panel of Figure \ref{fig:phase-lines}, which show single-peaked emission with no variation of its center wavelength.

On the other hand, the Doppler map of He II 4684 line shows a ring-like structure with two bright spots in the upper left ($V_{\rm x} \sim - 250$ km/s and $V_{\rm y} \sim 200$ km/s)  and right  ($V_{\rm x} \sim 250$ km/s and $V_{\rm y} \sim 100$ km/s) quadrant (lower panel of Figure \ref{fig:dopmap}).
The bright annuli has the velocity (200-400 km/s) significantly lower than the rotation velocity at the tidal truncation radius under Kepler rotation.
Therefore, He II 4686 \AA~  might not originate from the accretion disk as well as H$\alpha$.
We note that, as we described in Section \ref{sec:3}, the H$\alpha$ and He II 4686 emission line can be blended with  Pickering series 6560 \AA~ and Bowen blend, respectively.
However, a blended line only creates a ring-like structure \citep{mar16dopmapbook, neu20HTCasoutburstdopmap}, and hence we simply ignored the contribution from these lines.

\section{Discussion}
\label{sec:5}

\subsection{Doppler tomography of H$\alpha$: origin of single-peaked emission}
\label{sec:5.1}

In  Doppler tomography of H$\alpha$ (upper panel of Figure \ref{fig:dopmap}), V455 And showed a compact blob centered at the primary WD.
On the other hand, that of WZ Sge, which has very similar binary parameters as V455 And \citep{sma93wzsge, ara05v455and}, showed almost flat emission from the accretion disk in H$\alpha$ during the early superhump phase \citep{bab02wzsgeletter}.
This fact is consistent that WZ Sge showed a double-peaked  emission of H$\alpha$, whereas V455 And showed a single-peaked emission profile during the early superhump phase.

This remarkable compact blob observed in V455 And clearly does  not originate from the accretion disk, since the accretion disk is projected as a ring structure in Doppler map \citep{DopplerTomography}.
The line profile of H$\alpha$ is stable across the orbital phase even during the eclipse (left panel of Figure \ref{fig:phase-lines}), suggesting that the emitting region of H$\alpha$ is above the orbital plane.
A similar compact blob centered at the primary WD in  Doppler maps is observed in some SW Sex-type CVs (e.g., \cite{hel96v1315aql, rod01v348pup, her17rwsexj0644} ).
SW Sex-type CVs are classically defined as an eclipsing nova-like system showing single-peaked emission lines \citep{tho91pxand}.
Nowadays they include some non-eclipsing objects sharing similar spectroscopic properties (e.g.,  \cite{rod07newswsex}).
Since a DN around  outburst peak has a high mass-accretion rate comparable to nova-like systems and SW Sex-type CVs,
the disk condition and the projected observational properties of eclipsing V455 And can be similar to SW Sex-type CVs, while the mass transfer rate is significantly different.
In SW Sex-type CVs, the compact blob centered on the primary WD is the most prominent in high-excitation lines such as He II 4686 \AA.
The Doppler maps in Balmer lines of SW Sex-type CVs are, on the other hand, usually dominated by an emission from the accretion disk and the superimposed feature on the lower left quadrant, which is regarded as an emission from the re-impact region of the mass stream overflow \citep{hoa93CVwind, hel96v1315aql}.
In case of V455 And, as WZ Sge-type DNe are known as  CVs with the lowest mass-transfer rate (e.g., \cite{kni11CVdonor}) and as Doppler map does not show the evidence of the mass transfer burst, these re-impact region should not be formed even during the peak of a superoutburst and single-peaked component was not obscured even in H$\alpha$.

Even though the compact blob observed in V455 And can be an analogy of that in SW Sex-type CVs, the origin of this feature in SW Sex-type CVs is still not settled \citep{war95book}.
One possibility is the accreting material along the magnetic field of the primary WD.
Some of SW Sex-type CVs are known as intermediate polars through optical, X-ray and polarimetric studies (e.g., \cite{lim21SWSexasIP} and references therein).
\citet{wil89CVeclipse} proposed that the single-peaked emission lines in SW Sex-type CVs can  attribute to  the magnetic accretion curtain around the primary WD.
As the accreting materials follow the magnetic field lines above the disk plane, little eclipses of emission lines are observed \citep{wil89CVeclipse, cas96v795her, hoa03dwuma}.
V455 And is also known as an intermediate polar \citep{ara05v455and, sil12v455andGALEX, szk13V455And, gha20v455and}, hence the accretion along the magnetic fields can take place and a similar spectrum can be observed in V455 And.
We note, in the ordinary superhump phase of V455 And superoutburst Balmer lines exhibited a double-peaked profile.
Later phase of the outburst means the lower mass-accretion rate as well as the fainter the accretion disk itself is, hence the magnetic accretion could be more prominent in the ordinary superhump phase than during the early superhump phase.
This respect suggests that the magnetic accretion column might not be responsible for the single-peaked emission line of H$\alpha$ during the early superhump phase.

Another possibility is the disk wind component.
\citet{hon86swsex, hoa94CVHeII, hel96v1315aql, mur97diskwind} interpret the stable single-peaked emission as a wind component.
As calculated by \citet{hon86swsex, mat15CVdiskwind}, considering the wind component other than the accretion disk generally makes the line profile much narrower,
while the exact emission line profile depends on the applied model of the disk wind.
Since the wind should locate above the orbital plane, they would not be eclipsed so much.
However, the presence of an outflow in  DN outbursts is still not established.
\citet{kaf04windfromCV} reported optical P Cygni profiles in He I lines (5876 \AA~ and 7065 \AA) and H$\alpha$ of some nova-like stars, indicating the presence of an outflow in CVs.
More recently, radio detections during some DN outbursts \citep{kor08SSCyginradio, cop16DNinRadio} gives a hint of a transient jet in an early stage of DN outbursts, while the mechanism of radio emission in CVs is still not fully understood.
Even though the presence of wind or outflow has not been confirmed during the DN outburst in optical observations so far, combining the discussion on the origin of He II 4686 \AA, we propose that the emission from the wind components can be an alternative interpretation of the compact blob in Doppler map of H$\alpha$.

Even if this single-peaked emission  originated either from a magnetic accretion column or a wind component, the reason WZ Sge and more generally other eclipsing DNe in outburst did not show a similar feature is still controversial.
The face-on WZ Sge-type DN GW Lib showed a strong narrow emission in H$\alpha$ during its early superhump phase \citep{hir09gwlib}.
Since the emission component of GW Lib did not show any blue-shift, H$\alpha$ emission should not likely originate from an outflow in the case of GW Lib.
Therefore, the emission components observed in V455 And and GW Lib may have a different origin.
More other sample of time-resolved spectrum during the early superhump phase of WZ Sge-type DNe, as well as multi-wavelength observations to seek the impact of an outflow are needed to test the diversity of DN outbursts.

\subsection{Doppler map of He II 4686 \AA: not originate from the accretion disk}
\label{sec:5.2}

Since the discovery of the spiral structure in Doppler map of WZ Sge during the early superhump phase by \citet{bab02wzsgeletter, kuu02wzsge}, He II 4686 \AA~ emission line is regarded as a tracer of the spiral arms emerging from the 2:1 resonance and causing early superhumps.
Recently, \citet{tam21seimeiCVspec} studied both the optical spectra taken during the early superhump phase and the time-resolved photometory data of WZ Sge-type DNe.
They showed that the strength of He II 4686 \AA~ emission line is  positively correlated with  the amplitude of early superhump amplitude, therefore likely with the inclination of the system.
This result is consistent with the view that the spiral arms are responsible for both the early superhumps and the strong emission of He II 4686 \AA~ in WZ Sge-type DN outbursts.

The Doppler tomography of He II 4686 \AA~ in both V455 And and WZ Sge shows a ring-like structure plus the two flaring patterns.
The phases of the flaring spots are slightly different between WZ Sge and V455 And \citep{bab02wzsgeletter}.
The upper left spot in V455 And is shifted to the left quadrant ($V_{x} < 0$ and $V_y \sim 0$) in WZ Sge, although the right spot locates in almost same phase  ($V_{x} > 0$ and $V_y \sim 0$) in V455 And and WZ Sge.
Along with that, in the case of WZ Sge, the right spot is brighter than the left spot.
These differences of the flaring spots might reflect the exact structure of the spiral arms depending on the mass ratio of the systems, as well as the observation date since the outburst peak.
More interestingly,  the velocity of the ring structure is significantly different between WZ Sge and V455 And.
The ring structure of V455 And has the velocity significantly lower than 400 km/s, even though the one of WZ Sge has 500-600 km/s, which is consistent with Kepler velocity around the tidal truncation radius \citep{pac77ADmodel}.
Therefore, the emitting region of He II 4686 \AA~ in V455 And should differ from the Keplerian accretion disk in WZ Sge.

We first checked if the emission of He II 4686 \AA~ originates from the accretion disk, by considering a imaginary force other than centrifugal force to balance the gravitational force and to acquire the observed velocity of the accretion disk.
We measured the deviation from a Keplerian disk in the form of Equation \ref{eq:1}, where $G$, $M_{\rm WD}$, $R$, $\Omega$, and $f$ are the gravitational constant, a primary WD mass, disk radius, Keplerian orbital frequency, and fraction of a imaginary force, respectively.
The inclination is assumed to be  75$^\circ$ \citep{ara05v455and}.

\begin{equation}
\label{eq:1}
\frac{G M_{\rm WD}}{R^2}(1-f) = R \Omega^2
\end{equation}

where $f = 0$ corresponds to the ideal Kepler rotation.
Even if we assume that all the emission of He II 4686 \AA~ originate from $R = 0.6a$, corresponding to the tidal truncation radius, the ideal Keplerian rotation around the primary WD condition ($f = 0$) gives the emission with the velocity of $\sim 600$ km/s, which is too large for the observed value ($\sim 300$ km/s) in V455 And.
In fact, assuming Kepler rotation around the primary WD, rotating at 300 km/s corresponds to a radius bigger than twice of the binary separation.
It is not expected the disk to expand as big as twice the binary separation, because, as described by  \citet{osa02wzsgehump}, the angular momentum stored in the quiescence disk is not much enough to expand the disk beyond  $(7/5)^2R_{\rm cir}$, where $R_{\rm cir}$ is circularization radius and $R_{\rm cir} \sim 0.25a$ in $q=0.080$ \citep{lub75AD, hes90gd552}.
To obtain the same velocity as observed in V455 And, $f > 0.6$ is required.
Since the disk instability model in DN outbursts (e.g., \cite{mey81DNoutburst}) and $\alpha$ disk model \citep{sha73alphadisk} assume the Keplerian rotation in the accretion disk, these models would not work in the disk with such a large deviation from Keplerian disk.
Therefore, the ring-like structure of V455 And does not likely originate from the accretion disk.

We note that the spiral structure in the accretion disk is not only observed in WZ Sge-type DNe, but also in some U Gem-type DNe (e.g. \cite{har99ippeg, ste01spiralwave, gro01ugemspiral}).
However, the phase and structure of spiral waves seen in U Gem-type DNe and WZ Sge -type DNe are significantly different, suggesting that the origin of the spiral waves in these two types of DNe would be different.

\subsection{Comparison with the height map of the accretion disk}
\label{sec:5.3}

In order to further examine the origin of the peculiarity of He II 4686 \AA~ emission in V455 And, our Doppler map was compared with the disk height map obtained during the early superhump of V455 And in \citet{uem12ESHrecon}.
\citet{uem12ESHrecon} is the first case in which the height structure of the accretion disk during the early superhump phase was successfully reconstructed.
They revealed two flaring patterns in the outermost part of the disk along with the elongated structure into the inner disk.
In this context, this is the first case that, in a single object, we can directly compare both the height structure of the accretion disk and  Doppler tomography.
We note our data were not obtained on the same night as \citet{uem12ESHrecon}.
We mainly compared our results with the reconstructed map using the data on BJD 2454352 (Day 3 in \cite{uem12ESHrecon}), which is 2 days after our observations with HDS.
The flaring regions at the upper right quadrant ($V_{\rm x} \sim 250$ km/s and $V_{\rm y} \sim 200$ km/s) and the right quadrant ($V_{\rm x} \sim 250$ km/s and $V_{\rm y} \sim 100$ km/s) in  Doppler tomography  correspond to the upper left part and bottom part of the accretion disk in the height map of \citet{uem12ESHrecon}, respectively.
In these phases of the accretion disk, the inner arm structures which elongated relatively inside of the disk are located, suggesting a relation between He II 4686 \AA~ emission and the spiral arm structure.

Even though He II 4686 \AA~ was not dominated by the emission  from the accretion disk, this coincidence should be attributed to a type of asymmetry in the system.
Considering the fact that the profile of the emission lines was not affect by the eclipses and the peak separation of He II 4686 \AA~ was too small for the emission from the accretion disk, 
we propose that the wind component can be the origin of this peculiar He II 4686  \AA~ emission line during the early superhump stage of V455 And.
The peak separation of a few hundred km/s is expected in the emission line from a wind component \citep{mat15CVdiskwind}.
In this context, the asymmetry in the Doppler tomography is understood as the difference of the line emissivity on the phase of the wind, which could be related to the disk phase dependence of the mass flow rate of the wind.
Since the local escape velocity ($V_{\rm escape} = \sqrt{\frac{2GM_{\rm WD}}{R}}$) is faster than the local Kepler rotation velocity ($V_{\rm Kepler} = \sqrt{\frac{GM_{\rm WD}}{R}}$), and  the conservation of angular momentum of the wind results in the slower rotational velocity at outer radius along the wind trajectory, the little difference of phases of the flaring spot in the disk height map and the Doppler tomography can be understood.

We note that the reconstructed disk height map of OT J012059.6+325545 during the early superhump does not show the inner part of the spiral arms \citep{nak13j0120}.
While the spectrum of  OT J012059.6+325545 was not reported in literature, the uniqueness of V455 And can be understood as the presence of the inner spiral arms and resulting the disk wind emission.
We would need more samples of both the early superhump light curves and time-resolved spectroscopic observations to verify the nature of the wind in DNe outbursts.

As discussed in Section \ref{sec:5.1}, the single-peaked emission of H$\alpha$ also possibly arose from the wind component.
In SW Sex-type CVs, a single-peaked profile is usually observed in He II 4686 \AA, while the double peaked profile was observed in V455 And.
This is likely because our data was obtained with HDS mounted on Subaru telescope with very high wavelength and time resolution enough to resolve a more compact structure in the emission line variation.
By changing the total variation parameter of the Doppler tomography in \citet{uem15DTTVM}, more featureless map with a compact blob centered on the primary WD is generated even in He II 4686 \AA.
Since the excitation level of H$\alpha$ is much lower than He II 4686 \AA, 
H$\alpha$ line is emitted from more above the disk plane where the wind might reach slower velocity and hence show a narrower emission line.
Therefore, both H$\alpha$ and He II 4686 \AA~ emission lines can be simultaneously emitted from the disk wind component with different peak separations.

\subsection{Wind model}
\label{sec:5.4}

To verify whether the wind component can explain the narrow peak separation and non-axisymmetricity in He II 4686 \AA~ emission line, we constructed a simple model of an accretion disk wind.
In our model, we more focused on to reproduce the emission line with the peak separation of $< 500$  km/s and the variation of the emission profile across the orbital phase applying the non axisymmetric structure of the wind.
In our model, we assumed that the wind component is launched from the inner part of the accretion disk ($0.05a < R < 0.35a$).
This radius corresponds to the inner arm structure reconstructed in \citet{uem12ESHrecon}.
The initial velocity is 1.1 times of the local escape velocity (typically 1,000 km s$^{-1}$) from the primary WD to the radial direction and then the wind is decelerated by the gravitational force of the primary WD.
As the emission line of  He II 4686 \AA~ showed double-peaked profile across all the phases, the symmetric components should be presented in the wind component.
The non-rotational velocity field of the wind $v_{\rm l}$ is given as Equation \ref{eq:escapeV},

\begin{equation}
    \label{eq:escapeV}
    v_{\rm l} (R) = 1.1\sqrt{\frac{2 G M_{\rm WD}}{R_0}} 
        - \sqrt{2 G M_{\rm WD}  \left(\frac{1}{R_0} - \frac{1}{R}\right)},
\end{equation}

where $R_0$ is the launch radius of the wind.
We note that the factor of 1.1 is a marginal value, and we chose this value to best reproduce the observed peak separation of He II 4686 \AA.
At the same time, the rotational velocity field of the wind $v_\phi$ is obtained assuming  conservation of specific angular momentum (Equation \ref{eq:phiv}; \cite{mat15CVdiskwind}).

\begin{equation}
\label{eq:phiv}
    v_\phi(R) = \sqrt{\frac{G M_{\rm WD}}{R_0}} \frac{R_0}{R}
\end{equation}

As discussed in the previous section, the phases in which two flaring spots are observed in Doppler map correspond to those of the inner spiral arm structure in the disk height map.
We applied a non-axisymmetric structure in the wind by considering the additional emission from  the wind originally launched from (x,y) = ($-$0.10,$-$0.20) and (x,y) = (0.15, 0.25), corresponding to the inner arm structure on Day 3 in \citet{uem12ESHrecon}.
More specifically, the size of the inner arm structure is modeled with two-dimensional Gaussian (FWHM=0.1a), and at the peak of Gaussian the emission is twice enhanced compared to the other part of the disk. 
Since considering the physical structure of the disk wind and the detailed emission mechanism of each line is beyond the scope of this paper, we assumed that the emissivity decreases with the proportion of $1/R^2$ and summed all the wind components on the velocity toward the observer with the inclination of 75$^\circ$ \citep{ara05v455and}.

\begin{figure}[tbp]
 \begin{center}
    \includegraphics[width=65mm]{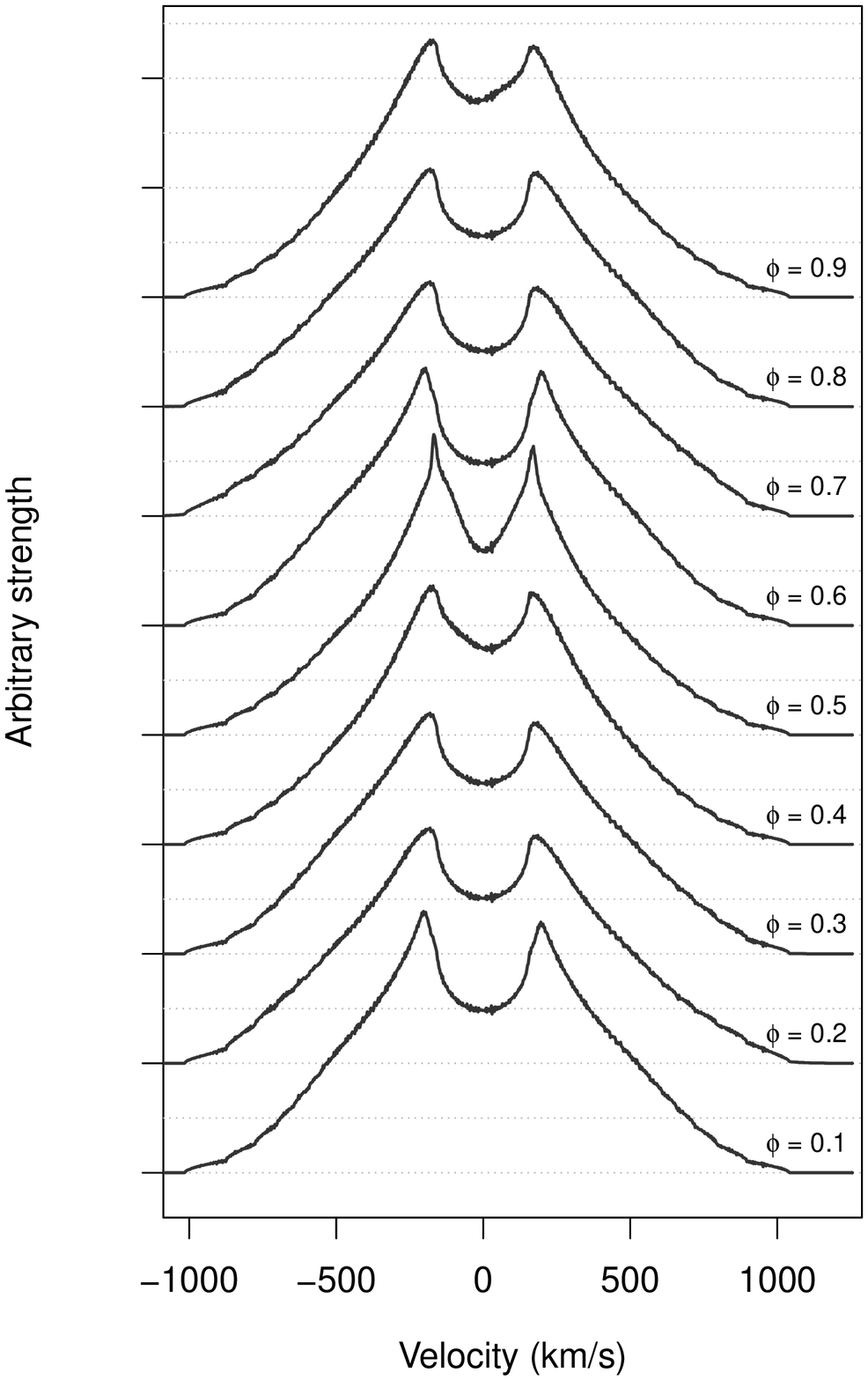}
    \includegraphics[width=83mm]{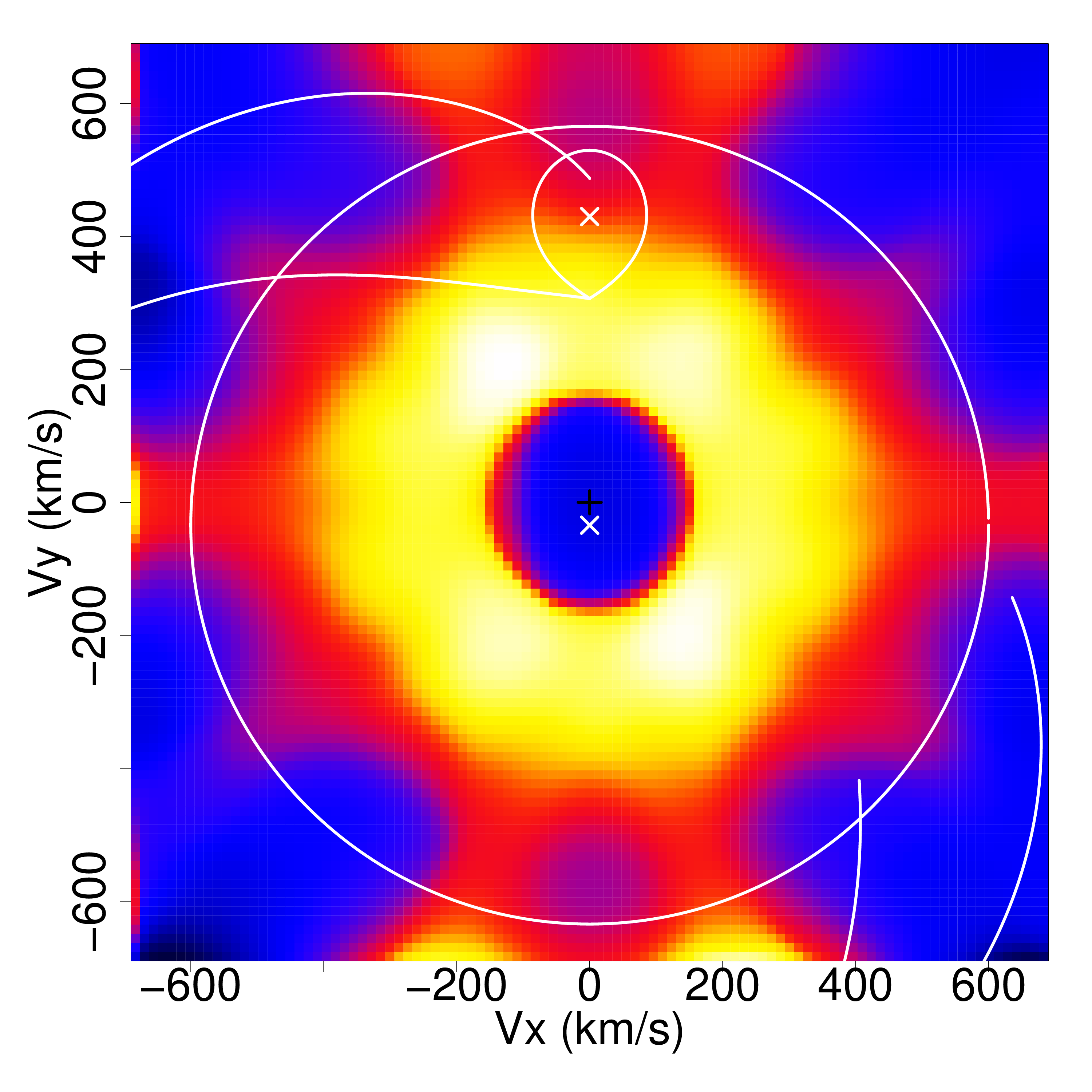}
  \end{center}
 \caption{
 Synthesised emission line from the wind component (left)
 and the Doppler map (right).
 The system values for the Doppler map is same as Figure \ref{fig:dopmap}
 }
 \label{fig:windmodel}
\end{figure}

In Figure \ref{fig:windmodel}, the synthesised emission line profile (left panel) and  Doppler map (right panel)  using our model are presented.
The peak separation of the emission profiles is $\sim 500$ km/s and observed in all orbital phases, consistent with our observations of He II 4686 \AA~ presented in the right panel of Figure \ref{fig:phase-lines}. 
In  Doppler map, two flaring spots are seen, which locate in the upper left quadrant ($V_x < 0$ and $V_y > 0$ ) and lower left quadrant  ($V_x > 0$ and $V_y < 0$).
The phase of flaring spots are slightly different from our observation.
This is likely due to the dependence of the launching speed of the wind ($V_0$) and more importantly, the lack of modeling the physical mechanism of the disk wind and radiative transfer.
This difference of flaring spot phases also might attribute the difference of the spiral arm structure to the date of spectroscopic observation and photometric observation.
Even with these respects, general properties are well reproduced, inferring that this narrow emission line of He II 4686 \AA~ is attributed to the accretion disk wind with two flaring spots.
More detailed analyses applying radiative transfer calculation and physical structure of the wind are required to confirm our results.
It is worth noting that the FWZI of our model emission profile is $\sim 2,000$ km/s, which is smaller than the observed value of $\sim 4,000$ km/s.
This is due to the fact that our model does not include the emission from the accretion disk, and therefore the observed broader profile should be accounted for the emission from the inner disk.

\subsection{The time evolution of spectra} 
\label{sec:5.5}

\begin{figure*}[tbp]
 \begin{center}
    \includegraphics[width=54mm]{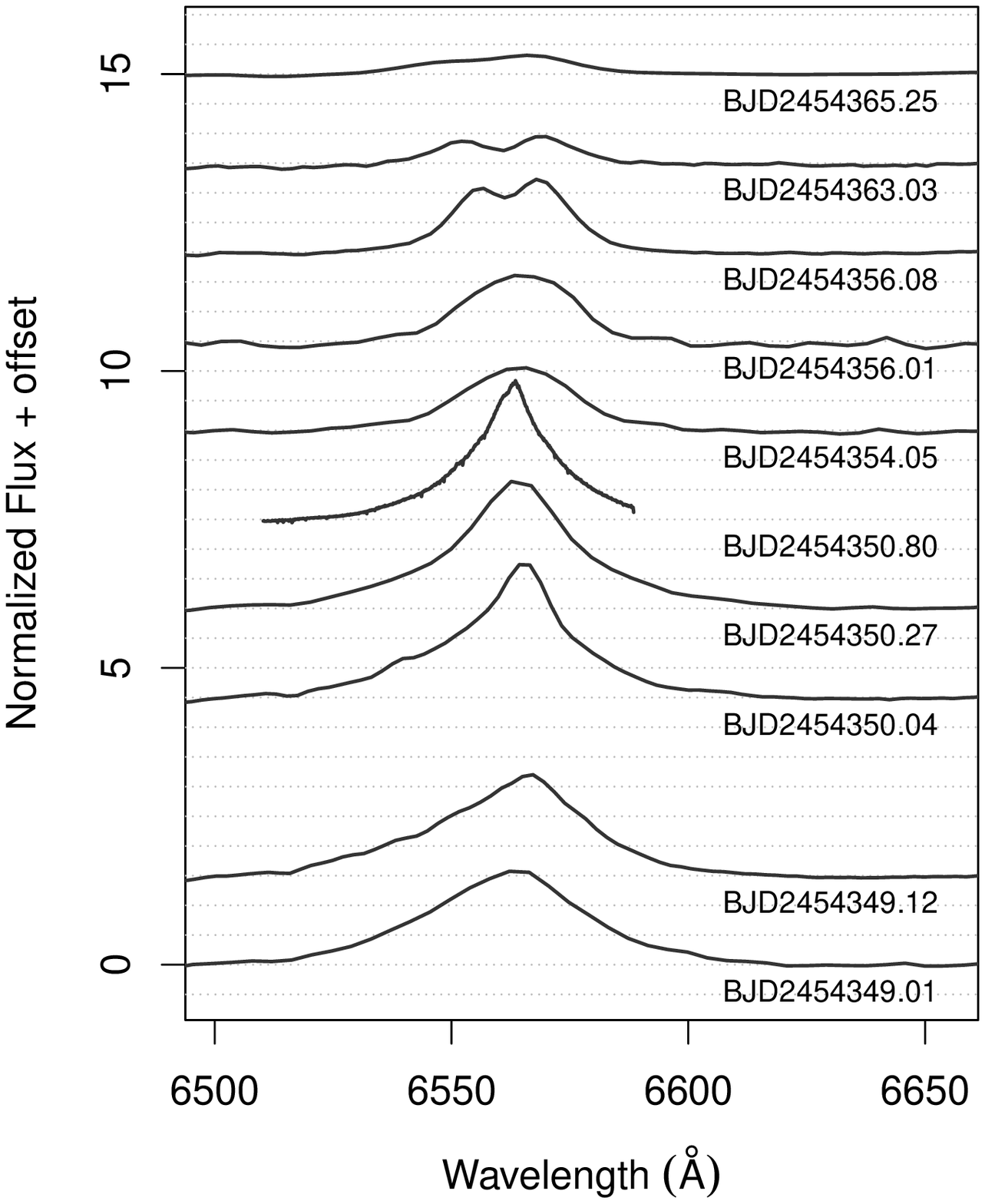}
    \includegraphics[width=54mm]{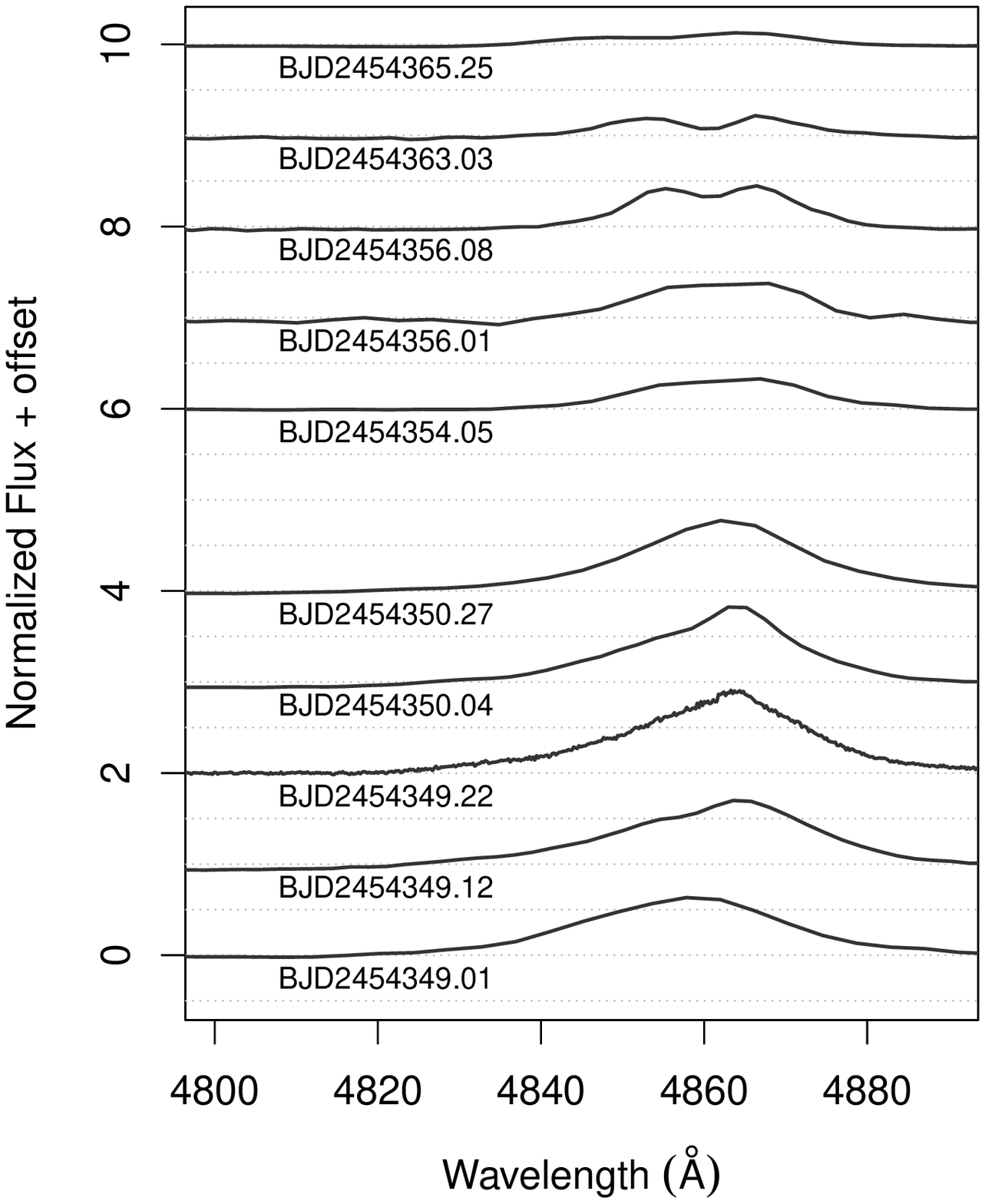}
    \includegraphics[width=54mm]{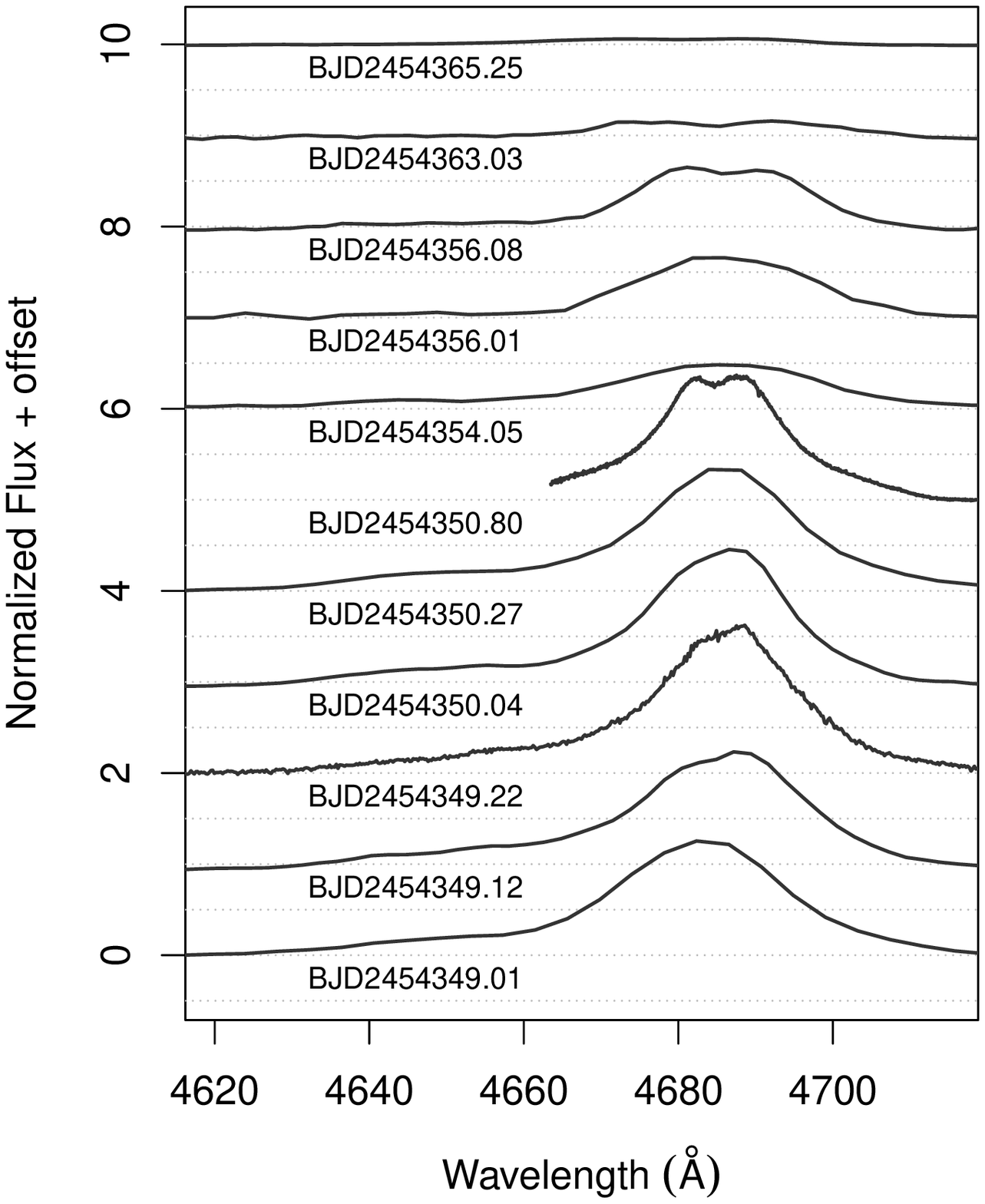}
  \end{center}
 \caption{
 Zoomed spectra around the H$\alpha$ (left), H$\beta$ (middle), and He II 4686  \AA~ (right) during the 2007 superoutburst.}
 \label{fig:spec-lines}
\end{figure*}

As described in Section \ref{sec:3}, V455 And showed a significant evolution of its spectrum throughout the 2007 superoutburst and post-outburst stage.
Figure \ref{fig:spec-lines} presents the zoomed spectral evolution of H$\alpha$ (left), H$\beta$ (middle), and He II 4686  \AA~ (right) lines during the 2007 superoutburst.
In this subsection, we more focused on these spectral evolutions compared with WZ Sge \citep{bab02wzsgeletter, kuu02wzsge,nog04wzsgespec} and GW Lib \citep{hir09gwlib}.
WZ Sge is a proto-type object of WZ Sge-type DNe, and WZ Sge and V455 And are thought to share similar binary parameters (e.g., short orbital period near the period minimum, high inclination showing grazing eclipse: \cite{sma93wzsge,ara05v455and}).
GW Lib is, on the other hand, a well-known low inclination WZ Sge-type DN which does not show detectable early superhumps \citep{tho02gwlibv844herdiuma, Pdot}. 
All WZ Sge, GW Lib and V455 And are spectroscopically observed during both the early superhump phase and ordinary superhump phase.

%in early superhump phase
During the early superhump phase after the maximum light, V455 And showed double-peaked He II 4686  and single-peaked Balmer line emission lines.
WZ Sge during the early superhump phase, on the other hand, showed double-peaked emission lines of He II 4686 \AA~ and H$\alpha$, while other Balmer series are in absorption \citep{nog04wzsgespec}.
This single-peaked Balmer lines in V455 And should be attributed to the component other than an rotating accretion disk as described in Section \ref{sec:5.1}, while the double-peaked  profile of WZ Sge should be understood as the emission from the accretion disk.

High excitation lines such as Bowen blend, C IV / N IV blend, Pickering He II 5412 \AA~ emission lines were detected in both V455 And and WZ Sge during the early superhump phase, while Pickering He II 4200 \AA~ and  4542 \AA~ were detected only in V455 And.
Former high excitation lines are indeed detected in various DN outbursts including WZ Sge-type DNe (\cite{tam21seimeiCVspec} and the reference therein). 
Contrast to that, stronger Pickering series is rarely observed in DN outburst and those in V455 And can be related to its nature as a intermediate polar \citep{har99J0558}.
Concerning the emission line strength, the EWs of V455 And is about order of magnitude stronger than those of WZ Sge, and the peak separation of He II 4686 \AA~ in V455 And ($\sim 375$ km/s) is much narrower than those in WZ Sge ($\sim 1,000$ km/s; \cite{bab02wzsgeletter,nog04wzsgespec}) as discussed in Section \ref{sec:5.2}.
Therefore, the emission lines of V455 And other than H$\alpha$ and He II 4686 \AA~ were also affected by the possible wind components.
These respects tell that even though WZ Sge and V455 And have very similar binary parameters, the system structure and conditions during the early superhump phase of the superoutburst were significantly different.
These differences between V455 And and WZ Sge should attribute to the nature of V455 And as a intermediate polar, and the possible  presence of an outflow component as discussed in previous sections.

In the case of GW Lib, He II 4686 \AA~ and H$\alpha$ were observed as single-peaked emission lines and the higher Balmer series are observed as an absorption line accompanying single-peaked weak emission core.
Even though the single-peaked emission profiles are consistent with its low inclination, face-on systems should show stronger absorption arose from the face-on optically thick accretion disk during the outburst \citep{mar90ippeg}.
In fact, even WZ Sge-type DNe with detectable low-amplitude early superhumps, which should have  larger inclinations than GW Lib, show H$\alpha$ line in absorption during the early superhump phase \citep{tam21seimeiCVspec}.
Therefore, the emission component of Balmer series in GW Lib could originate from other than the accretion disk, and may have a similar mechanism as V455 And.
We note that again, since no blue-shifts of emission lines are observed in GW Lib \citep{hir09gwlib}, this additional component observed in GW Lib might not be  an outflow.
To summarize, the larger variety of spectra was observed even among these three systems in the early superhump phase, which is highlighted by the spiral structure and hence a high accretion rate, suggesting a more complex disk and system structure during the early superhump phase in WZ Sge-type DN outbursts.

%in ordinary superhump phase
In the ordinary superhump phase, V455 And and WZ Sge showed similar strength and species of the lines \citep{nog04wzsgespec} in the viewpoints of EWs and peak separations.
H$\alpha$, He I 5876 \AA, 6678 \AA~ and He II 4686 \AA~ have double-peaked  emission profile in WZ Sge and V455 And, and both showed He I 4387 \AA, 4471 \AA, and 4912 \AA~ in absorption. 
Bowen blend was detected, while C IV / N IV were decayed out in both systems during the ordinary superhump phase.
Pickering series were not observed in this phase of V455 And as well.
In contrast to those lines, H$\beta$ was observed as emission in V455 And, while as absorption in WZ Sge.
The peak separations became larger from the early superhump phase to the ordinary superhump phase in V455 And and WZ Sge, suggesting that the decrease of the accretion rate and high-energy irradiating photon.
The spectra of GW Lib in the ordinary superhump period showed absorption lines of Balmer and He I, with a weak emission core in H$\alpha$.
Such a similarity between V455 And and WZ Sge, and the cookie-cutter spectra of GW Lib as a low-inclination DN during the ordinary superhump phase suggests that the disk and system conditions were also much more similar in the ordinary superhump phase than those in the early superhump phase.

As discussed above, great spectral evolutions were happened in all three systems during the early - ordinary superhump phase transition, attributing a significant transformation of the disk and system structure between these phases \citep{osa02wzsgehump}.
The decays of high ionization lines are happened during the early-ordinary superhump phase transition and  either because of the decay of the mass accretion rate and the resulted decay of high energy photons to irradiate, and/or because of the disappearance of the spiral structure and resulted cooling of the accretion disk, as interpreted by \citet{nog04wzsgespec, hir09gwlib}.
As discussed in Section \ref{sec:5.4}, if the single-peaked emission of H$\alpha$ and narrow-separation He II 4686 \AA~ during the early superhump phase in V455 And were the evidence of the wind component,  the disk wind was present only during the early superhump phase.
While H$\alpha$ of GW Lib and V455 And was decayed after the early-ordinary superhump transition, one of WZ Sge strengthened.
According to \citet{hir09gwlib}, this difference can be understood as the presence of the mass reservoir, which are the outer gas of the accretion disk and will result in the mass donor for  rebrightenings \citep{kat98super}.
As V455 And also did not show any rebrightenings along with GW Lib (Type-D rebrightening; \cite{Pdot, ima06tss0222}), this interpretation still holds among V455 And, GW Lib and WZ Sge.

In the post-superoutburst stage, our spectra are resembled to the spectra observed in the quiescence before the 2007 superoutburst \citep{ara05v455and} and after the superoutburst \citep{szk13V455And} in the context of line series and profiles.
The EWs of the emission lines in our spectra are slightly weaker than those in quiescence, due to the brighter magnitude which attributes to the enhanced emission from the cooling primary WD and the accretion disk \citep{ara05v455and, szk13V455And}.
WZ Sge in the quiescence between rebrightenings and GW Lib in the post superoutburst stage as well showed the emission of Balmer and He I \citep{nog04wzsgespec, hir09gwlib}.

\section{Summary}
\label{sec:6}

We report on the spectroscopic observations of V455 And taken during its 2007 superoutburst and following post-superoutburst stages.
Our key findings are summarized as follows.

\begin{itemize}

\item                                            
During the early superhump stage, Balmer series showed a single-peaked emission profile, which is inconsistent with the high inclination of V455 And and did not likely originate from an accretion disk.
High excitation lines such as He II 4686 \AA, Bowen blend, C IV / N IV blend were observed, and those were also detected in the early superhump stage of WZ Sge.
He II 4686 \AA~ had a double-peaked profile with a peak separation of $\sim 440$ km/s, however this value is too low for emerging from an accretion disk during a DN superoutburst of V455 And.
Pickering series were observed during the early superhump stage as well.
As they are often observed in magnetic CVs, Pickering series can be related to the nature of V455 And as an intermediate polar.

\item
We performed Doppler tomography using the time-resolved spectroscopic data observed with HDS mounted on the Subaru telescope.
The Doppler map of H$\alpha$ showed a compact blob centered on the primary WD.
In analogy to SW Sex-type CVs, this feature can be arisen either from the disk wind component and/or the magnetic accretion column onto the primary WD.
However, it is a open question why other eclipsing DNe in outbursts do not show similar Doppler maps.

\item
Our Doppler map of He II 4686 \AA~ is dominated by a low velocity ring-like structure with two superimposed flaring regions.
The velocity of the ring-like structure is too slow to present the emission from an accretion disk with Kepler rotation.
The flaring regions in Doppler map share the same phase with the inner spiral arms formed by the 2:1 resonance, suggesting that the spiral arms are responsible for the non-axisymmetric emission of He II 4686 \AA, as well as early superhumps in the light curve.
One  possible scenario for such a narrow but double-peaked  emission line is to originate from a disk wind component.
Our simple model of an accretion disk wind with two non-axisymmetric flaring directions successfully reproduced the narrow peak separation and two flaring patterns on Doppler map.
This is the first case  among DN outbursts that the presence of an outflow component is inferred from an optical spectrum.

\item
While V455 And and WZ Sge are thought to share similar binary parameters, their system structures during the early superhump phase should be significantly different, especially in the respect of the magnetic activity of the primary WD and an additional component to produce the narrower strong emission lines in V455 And.

\item
During the ordinary superhump phase, V455 And exhibited Balmer series and He II lines in double-peaked  emission profile, whose peak separations can be explained by a Keplerian accretion disk.
Pickering series and C IV / N IV lines were not observed during this phase, suggesting that their origin is related to the high accretion rate during the early phase of the outburst.
In the viewpoint of line species and strength, V455 And and WZ Sge showed a very similar spectrum.
Therefore, the system and disk structure during the ordinary superhump phase are much similar than those in the early superhump phase.

\item
The remarkable transitions of optical spectra in WZ Sge and GW Lib, and more dramatically in V455 And have happened in the early-ordinary superhump transition.
This spectral evolution between superhump stages should attribute to the evolution of the disk condition during the superoutburst of a WZ Sge-type DN, as predicted in the thermal-tidal disk instability model.

\end{itemize}

%%%%%%% acknowledgement %%%%%%%%%%%%%
\begin{ack}

Y.T. acknowledges support from the Japan Society for the Promotion of Science (JSPS) KAKENHI Grant Number 21J22351.
T.K and D.N. acknowledge support from the Japan Society for the Promotion of Science (JSPS) KAKENHI   Grant Number 21K03616.
This work is partly supported by JSPS KAKENHI Grant Number JP18H05439, the Astrobiology Center of National Institutes of Natural Sciences (NINS) (Grant Number AB031010).
We acknowledge the staff members of NHAO.
Part of this research is based on data collected at Subaru Telescope,
which is operated by the National Astronomical Observatory of Japan.
We are honored and grateful for the opportunity of observing the 
Universe from Maunakea, which has the cultural, historical and natural 
significance in Hawaii.

\end{ack}

%%%%%%%%%%%%%%%%%%%%%%%%%%%%%%%%%%%%

%%%%%%%%%%%reference items%%%%%%%%%%%%%%%%%%%

\bibliographystyle{pasjtest1}
\bibliography{cvs}

%%%%%%%%%%%%%%%%%%%%%%%%%%%%%%%%%%%%%

\appendix

\end{document}